\documentclass[12pt]{iopart}

\usepackage{graphicx}
\usepackage{dcolumn}
\usepackage{bm}
\usepackage{hyperref}
\hypersetup{colorlinks=true, citecolor=blue, urlcolor=blue, linkcolor=blue}
\usepackage{color}
\usepackage{dsfont}
\usepackage{subfigure}
\usepackage{graphicx}
\usepackage{epsfig}
\usepackage{epstopdf}
\usepackage[english]{babel}
\usepackage[T1]{fontenc}
\usepackage[utf8]{inputenc}
\usepackage[normalem]{ulem}

\begin{document}

\title[ ]{Role of continuum in nuclear direct reactions with one-neutron halo nuclei: a one-dimensional model}%

\author{Laura Moschini$^{1,*}$, Antonio M. Moro$^2$ and Andrea Vitturi$^{3,4}$}
\address{$^1$ Physics Department, University of Surrey, Guildford GU2 7XH, United Kingdom\\
$^2$ Departamento de F\'isica At\'omica, Molecular y Nuclear, Universidad de Sevilla, Apartado 1065, E-41080 Sevilla, Spain\\
$^3$ Dipartimento di Fisica e Astronomia 'G. Galilei', Università degli Studi di Padova, Via Marzolo 8,
35131 Padova, Italy\\
$^4$  INFN, Sezione di Padova, Padova, Italy}
\ead{$^*$l.moschini@surrey.ac.uk}

\begin{abstract}
We study the evolution of a single-particle wave function during the collision 
of a one dimensional potential well by another well, 
which can be regarded as a simple model for the problem of the scattering 
of a one-neutron halo nucleus by another nucleus. 
This constitutes an effective three-body problem, 
whose solution in three dimensions can be extremely complicated,
particularly when breakup and rearrangement channels are to be considered. 
Our one-dimensional model provides the essential three-body nature of this problem,
and allows for a much simpler application and assessment of different methods of solution.  
To simplify further the problem, we assume that the potential well representing the projectile 
moves according to a predetermined classical trajectory, 
although the internal motion of the ``valence'' particle is treated fully quantum-mechanically. 
This corresponds to a semiclassical approach of the scattering problem.
Different approaches are investigated to understand the dynamics involving one-body halo-like systems:
the ``exact'' time-dependent solution of the Schr\"odinger equation
 is compared to a numerical continuum-discretized coupled-channels (CC) calculation
presenting various model cases including different reaction channels.
This framework allows us to discuss the reaction mechanism and the role of continuum, whose inclusion in the CC calculation results to be crucial to reproduce the ``exact'' solution, even when the initial and final states are well bound. We also link each dynamical situation with analogous problem solved in a three dimensional (3D) CC framework, discussing the main challenges experienced in the usual 3D models.
\end{abstract}

\section{Introduction}
       One of the most relevant research lines in nowadays nuclear physics is the investigation, both
       experimentally and theoretically, of nuclei under extreme conditions and, 
       in particular, nuclei far from the stability valley.
       Along this line, truly enticing and sometimes striking novel nuclear structure phenomena are being observed.
       In particular, these so-called exotic nuclei present low proton or neutron separation energy, short lifetime, 
       and a radius which noticeably deviates from the $A^{1/3}$ dependence of stable nuclei,
       a fact that is related to a skin or a halo structure \cite{Tanihata3,Tanihata4}.
       The excess of neutrons or protons also leads to a different arrangement
       which is generally described as deformation: this is indicated by
        new single-particle character of states and new magic numbers related to new shell closure
        \cite{Nilsson1,Nilsson2}.
       Because of the large spatial separation between the centre-of-mass and the centre-of-charge,
       in presence of an external electric field 
       low energy electric dipole excitations can result, as well as very clear cluster effects \cite{Heyde}.
       Examples of exotic systems are nuclei with large neutron excess, 
       with the barely bound outermost ones creating an extended density distribution, referred to as halo.
       There can be different kinds of halo systems: a nucleus with one valence neutron is for example $^{11}$Be, 
       whereas $^{8}$B is an isotope presenting a one-proton halo.
       
       Nowadays, measuring the properties of such nuclei is the goal 
       of the main experimental nuclear facilities around the world  \cite{Tanihata,Tanihata2}.
       Due to their short lifetimes targets of such nuclei are not feasible,
       so they can not be studied by usual spectroscopic techniques.
       Thus, the challenge to measure nuclei on the drip lines has triggered 
       the development of radioactive nuclear beams \cite{TANIHATA1995505}.
Most of our present knowledge of stable and exotic nuclei stems from the analysis of nuclear direct reactions.
Direct reactions are characterized by different channels: elastic and inelastic scattering,
transfer of nucleons between the colliding nuclei, breakup (i.e.\ excitation to positive energy states), 
and they will feed a particular channel in a way that depends sensitively on its character \cite{Satchler}.
In particular, inelastic scattering excites collective states strongly; 
one-neutron transfer probes the single-particle character of states; 
two-nucleon transfer goes preferentially to states that exhibit strong pairing correlations;
pairing could also be tested via multi-nucleon transfer, as well as clustering; 
the role of continuum is investigated through breakup reactions.

       From a theoretical point of view, the conventional theories valid for the description of stable nuclei
       have difficulties to describe exotic systems, thus indicating the necessity to improve them and include
       the description of the new phenomena taking place at the limits of stability,       
       taking into account the new features.
      The theoretical description of halo nuclei is strongly characterized by its weakly-bound nature: 
      the valence nucleons are so weakly-bound that the addition of any correlation 
       to the simple mean field model inevitably 
       involves the inclusion of the continuum in the system description.
       For this reason the description of their structure or dynamics is more
       involved, even considering inert cores. 
Direct reactions are usually studied applying coupled-channels calculations, in the case of processes involving exotic systems the continuum discretized coupled-channels (CDCC) description is applied \cite{CDCC1,AUSTERN1987125,thompson_nunes_2009,YOMM12}. However it is often difficult to model the continuum, e.g.\  when transfer or breakup channels are dominant \cite{cdcc-lim1,cdcc-lim2,cdcc-lim3}.  
In this work we construct a simple framework in which we can test coupled-channels (CC) techniques
 and directly compare the results to the direct solution of the time-dependent Schr\"odinger equation.
 
We consider the problem of one particle\footnote{Note that, 
since we do not simulate Coulomb barriers, we assume the valence particle to be a neutron.},
 initially bound in a one-dimensional potential well (the ``target''), 
subject to an external field, which is also represented by a potential well (the ``projectile''), 
and which moves in one dimension according to a predetermined classical equations of motion.
This situation describes schematically the collision between two nuclei, each one modeled by a potential well, 
and it permits to understand the mechanisms which govern a nuclear reaction 
by following the evolution of the wavefunction associated to the particle.
Initially, since we assume the particle to be a valence nucleon of the target,
it coincides with an eigenstate of the target potential well.
During the reaction the particle will feel the interaction with the incident nucleus, 
i.e.\ the projectile potential well.
Depending on the parameters characterizing the reaction, 
the particle will more likely remain in its initial state (elastic scattering),
jump to an excited bound level of the target (inelastic scattering), be
transferred to a bound level of the other nucleus,
or it could leave the initial nucleus and escape to the continuum (breakup). 
As a consequence, the particle wavefunction will change,
according to the probability to excite the different reaction channels.

This article is structured as follow: 
in Sec.~\ref{model} we detail the model for the numerical ``exact'' solution  (Sec.~\ref{exact})
as well as the coupled-channels description (Sec.~\ref{CC}). 
In particular, in  Sec.~\ref{continuum}, we discuss the treatment of continuum in our CC calculation.
In Sec.~\ref{results} we apply our models to different dynamical situations; 
finally we summarize our results and give our conclusions in Sec.~\ref{conclusion}.

\section{The model \label{model}}
Let us start the model description by specifying the Hamiltonian and the initial conditions.
The system is described with the Hamiltonian
\begin{equation}
  {\cal H}(x,t) = -\frac{\hbar^2}{2\mu}\frac{\partial ^2}{\partial x^2} + V_T(x) +  V_P(x-x_p(t),t),
  \label{Hamiltonian}
\end{equation}
where we include two potentials $V_T$ and $V_P$, chosen with a Woods-Saxons shape,
 and associated to target and to projectile, respectively.
In our ``semiclassical'' model the target is at rest and only the projectile
potential well moves according to a classical trajectory.
We use the trajectory proposed in \cite{Esbensen}
\begin{equation}
  x_p(t) = x_0 + \sqrt{\rho^2 + (vt)^2} - \rho,
  \label{esbensen_trajectory}
\end{equation}
that accounts for the projectile motion with fixed
asymptotic velocity $v$ at large distances, corresponding to an asymptotic energy $E_{P} = 1/2mv^2$,
which determines the effective duration of the reaction (the higher the asymptotic energy, the faster is the reaction),
and a distance of closest approach $x_0$ between the two centers of the potentials.
This trajectory allows the projectile to change its acceleration
over the distance $\rho$ (in our model of the order of $2$~fm), thus simulating the nuclear interaction with the target;
at $t = \pm \infty$ the trajectory tends to a uniform motion with zero acceleration.
We determine the time interval as $dt = dx/4v$, where $dx$ corresponds to the spatial grid step length.
The turning point of the collision corresponds to $t = 0$.

Upon variation of these parameters, one can simulate different kinematical
conditions due to different bombarding energies and impact parameters 
(corresponding to partial waves in a quantum-mechanical description), 
while the choice of the parameters of the two potentials accounts for
the different masses of the colliding nuclei, the Q-values of the different
final channels as well as the possibility of simulating weak binding conditions.

By solving separately  the time-independent Schr\"odinger equation for each well 
\begin{equation}
   {\cal H}_J \Phi_J(x)  = E_J \Phi_J(x),
  \label{Schr\"odinger_equation_time_independent}
\end{equation}
with the Hamiltonian
\begin{equation}
  {\cal H}_J = \left[ -\frac{\hbar^2}{2\mu}\frac{d^2}{dx^2} + V_J(x) \right],
  \label{hamilt}
\end{equation}
we obtain two sets of bound levels $\Phi_J(x)$ associated to negative eigenvalues\footnote{Bound states are obtained by solving 
the finite-difference method on the whole grid \cite{numerov}.}, 
and a continuum associated to positive energies.
By applying the discretization methods detailed in Ref.~\cite{MPV2016}, 
we can also define a set of square-integrable functions associated to a finite number of positive energies,
the so-called discretized continuum pseudostates.

Depending to the kind of reaction under study
we will select one of the bound levels as the initial state of our single-particle
wave function. 
For example, to describe a pick-up reaction we will choose as initial state a
single-particle level in the target.
In the simulations presented here we will always choose a target wavefunction as initial one.

The problem can be solved in many different ways. Two such methods are considered and compared here.
On one hand, we consider the exact solution of the problem by numerically solving 
the corresponding time-dependent Schr\"odinger equation  (section \ref{exact}). 
On the other hand,  we  consider an approximate solution 
using the coupled-channels method, which is usually applied in solving three-dimensional scattering problems  
(sections \ref{CC}). From this comparison, we expect to get further insight on the accuracy and limitations of the coupled-channels approach to quantum collisions.

\subsection{Exact time-dependent solution}\label{exact}
In the case of the ``exact'' solution, we proceed to compute the time evolution 
of the valence neutron wavefunction $\Psi(x,t)$ by
numerically solving the time-dependent Schr\"odinger equation
\begin{equation}
  i \hbar \frac{\partial}{\partial t} \Psi(x,t) = {\cal H}(x,t) \Psi(x,t),
  \label{Schr\"odinger_equation_time_dependent}
\end{equation}
with the Hamiltonian (\ref{Hamiltonian}).\\
The wavefunction $\Psi$ is confined within an interval containing the two wells.
It is calculated at fixed points with coordinates $x_{\mu}$ separated by $dx$.
According to \cite{Esbensen-ref7}, one can solve the problem by using a finite-difference approximation of the Hamiltonian, giving rise to the following  tridiagonal form:
\begin{equation}
        {\cal H}_{\mu \nu} = - \frac{\hbar ^2}{2mdx^2} \left( \delta_{\mu \nu +1} + \delta_{\mu \nu -1} -2\delta_{\mu \nu} \right)
        +\delta_{\mu \nu} \left[V_T(x_{\mu}) +V_P(x_{\mu}-x_p(t))\right],
  \label{Hamiltonian_tridiagonal}
\end{equation}
and the time evolution of the wavefunction is governed by the so-called Pad\'e approximation of the evolution operator:
\begin{equation}
        \Psi(t+dt) = \left(1+\frac{i dt}{2 \hbar}{\cal H}\right)^{-1}\left(1-\frac{i dt}{2 \hbar}{\cal H}\right)\Psi(t),
  \label{pade-approx}
\end{equation}
where $dt$ is a finite time step, and ${\cal H}$ is the matrix in equation (\ref{Hamiltonian_tridiagonal})
at the intermediate time $t+dt/2$. Note that the evolution operator is a unitary operator.\\
An alternative approach is to integrate the differential equation using a finite-difference method, 
such as the Runge-Kutta method. 
For that, in this work we make use of the routines  \textit{D02PVF} and \textit{D02PCF} of the \textsc{nag} library\footnote{We 
are also imposing vanishing boundary conditions.}. 
Although this solution prevents us from a complete control of the code,
it was found to be faster than the Pad\'e method. 
We have also verified that both methods lead to identical results.

At the end of the time evolution, we can compute the final probabilities for
each reaction channel by projecting the final wavefunction $| \Psi(x, t_{f}) \rangle$ 
onto the corresponding eigenstates of each well
obtained by solving equation (\ref{hamilt}) for each potential
(target bound states $\Phi^T_{i}(x)$ for elastic and inelastic, projectile bound states $\Phi^P_{i}(x-x_p(t_f))$ for transfer channels)
\begin{eqnarray}
   {\cal P}_{elastic} = | \langle \Phi^T_{g.s.}(x) | \Psi(x, t_{f}) \rangle | ^2, \\
   {\cal P}_{inelastic} = | \langle \Phi^T_{i \neq g.s.}(x) | \Psi(x, t_{f}) \rangle | ^2, \\
   {\cal P}_{transfer} = | \langle \Phi^P_{i}(x-x_p(t_f)) | \Psi(x, t_{f}) \rangle | ^2.
  \label{el-in-tr_probl}
\end{eqnarray}
We can also evaluate the breakup probability either by direct subtraction
\begin{equation}
  {\cal P}_{breakup} = \mathds{1} - {\cal P}_{elastic} - {\cal P}_{inelastic} - {\cal P}_{transfer},
  \label{breakup_subtraction}
\end{equation}
or by projecting the final wavefunction onto a complete set of continuum states $\varphi(k,x)$ depending on the asymptotic wave number $k=\pm \sqrt{\frac{2\mu E}{\hbar^2}}$: 
\begin{equation}
  \label{breakup_continuum_overlap}
  {\cal P}_{breakup} =  \int dE \sqrt{\frac{\mu}{2E\hbar^2}} {\cal P}(k) =  \int dk |\langle \varphi(k,x)| \Psi(x, t_{f}) \rangle | ^2.
\end{equation}
 In 1D, for each positive energy there are two degenerate continuum wave functions with momentum $k$, one incoming from the left and the other from the right. For each energy we took the symmetric and antysimmetric combinations of the momentum normalized continuum wave functions \cite{MPV2016,Moschini_thesis:2017}.
 
\subsection{Approximate solution within the coupled-channels method}\label{CC}
The same problem can be solved with the so-called coupled-channels method, 
which is a popular framework used to describe quantum collision problems in atomic, molecular and nuclear physics.  
For this calculation we follow the formulation of reference \cite{Esbensen}, 
and we take into account two finite sets of wavefunctions, 
related to the target and the projectile potentials: 
$\Phi_j^T(x)$ and $\Phi_j^P(x)$, of $N_T$ and $N_P$ states respectively.
For collisions among tightly-bound systems, 
the basis expansion is usually restricted to bound states of the projectile and target systems. 
However, when one of them is weakly bound, it is important to include also continuum states. 
For that, it is convenient to use a discrete representation of square-integrable functions, 
as those discussed in \cite{MPV2016}
(we will discuss in more detail the inclusion of continuum 
in coupled-channels method in section \ref{continuum}).
Moreover, they are defined in a one-dimensional spatial grid, whose origin corresponds to the center 
of the target potential, which also corresponds to the laboratory frame.
A different choice, like the center of mass frame of reference in which the two potentials are moving,
would have implied a careful treatment of target and projectile wavefunctions due to the     
non-covariance of Schr\"odinger equation (see Appendix C of \cite{Moschini_thesis:2017}).
In addition, these two bases are non-orthogonal so we will solve this problem introducing the dual bases $\omega_j^{(T,P)}(x,t)$,
as explained in \cite{Esbensen,Moschini_thesis:2017,Esbensen-ref4,Esbensen-ref5,Esbensen-ref6}. 
They are respectively associated to each well and
conjugate to the channel wavefunctions of each potential, through the definition
\begin{equation}
  \langle \Psi_m^I | \omega_n^J \rangle = \delta_{I,J}\delta_{n,m},
  \label{dual_basis}
\end{equation}
where $ I,J = T, P$ and $n,m = 1, 2, ... N_{(T,P)}$.

In the CC approach, the wavefunction describing the entire system is expressed as a combination of target and projectile states
\begin{equation}
  \Psi(x, t) = \sum_{j=1}^{N_T} c_j^T(t)\Phi_j^T(x) +  \sum_{j=1}^{N_P} c_j^P(t)\Phi_j^P(x-x_p(t)),
    \label{wf_CC_combination}
\end{equation}
and the solution of the problem is reduced to the determination of the time
evolution of the coefficients $c_j^{(T,P)}(t)$ from the finite set of coupled differential equations 
\begin{eqnarray}
  i \hbar \frac{\partial c^T_j}{\partial t} = \sum c^T_k \langle \omega_j^T | V^P |\Phi_k^T \rangle                        
+ \sum c^P_k \langle \omega_j^T | V^T |\Phi_k^P \rangle, \nonumber\\ 
  i \hbar \frac{\partial c^P_j}{\partial t} = \sum c^T_k \langle \omega_j^P | V^P |\Phi_k^T \rangle                        
+ \sum c^P_k \langle \omega_j^P | V^T |\Phi_k^P \rangle .
  \label{coupled-channels}
\end{eqnarray} 
These equations are solved with the initial conditions $c^P_j(t = -\infty) = 0$ and $c^T_j(t = -\infty) = \delta _{i,j}$, 
where $i$ indicates one of the bound states in the target potential well.\\
To derive equations (\ref{coupled-channels}), we first insert the equation (\ref{wf_CC_combination}) 
of the total wavefunction $\Psi(x, t)$ into the time-dependent Schr\"odinger 
equation (\ref{Schr\"odinger_equation_time_dependent}), thus obtaining
\begin{equation}
  \sum_j i \hbar \frac{\partial c^T_j}{\partial t} \Phi_j^T
+ \sum_j i \hbar \frac{\partial c^P_j}{\partial t} \Phi_j^P = 
  \sum_j  c_j^T({\cal H}-{\cal H}_T)\Phi_j^T 
+ \sum_j  c_j^P({\cal H}-{\cal H}_P)\Phi_j^P
     \label{CC_1}
\end{equation}
where ${\cal H}_J$ is the Hamiltonian corresponding to the potential well $J = T,P$ of equation (\ref{hamilt}),
and ${\cal H}$ is the Hamiltonian of the full system presented in equation (\ref{Hamiltonian}).

In problems involving different mass partitions, 
one may use the so-called prior and post representations of the Hamiltonian, 
depending on whether one considers the projectile-target combination of the initial or final states;
by definition they have to give the same results.
Equations (\ref{wf_CC_combination}) and (\ref{coupled-channels}) are constructed in prior representation.
In post representation we can expand the system wavefunction on the dual basis
\begin{equation}
     \Psi(x, t) = \sum_n {\tilde c}^T_n \omega^T_n + \sum_n {\tilde c}^P_n \omega^P_n,
  \label{post1}
\end{equation}
and, following a derivation similar to the one given for the prior representation,
we obtain the corresponding set of  coupled-equations
\begin{eqnarray}
  i \hbar \frac{\partial {\tilde c}^T_n}{\partial t} = \sum {\tilde c}^T_m \langle \Phi_n^T  | V^P |\omega_m^T \rangle                        
                                                    + \sum {\tilde c}^P_m \langle  \Phi_n^T | V^P | \omega_m^P\rangle, \nonumber\\ 
  i \hbar \frac{\partial {\tilde c}^P_n}{\partial t} = \sum {\tilde c}^T_m \langle \Phi_n^P  | V^T |\omega_m^T \rangle                        
                                                    + \sum {\tilde c}^P_m \langle  \Phi_n^P | V^T | \omega_m^P\rangle .
  \label{coupled-channels_post}
\end{eqnarray} 
From the total wavefunction $\Psi(x, t)$, we can also extract amplitudes for excitation and transfer
in post and prior representations through the expressions 
\begin{eqnarray}
  {\tilde c}^I_n = \langle \omega_n^I  | \Psi \rangle,  \nonumber\\ 
          c^I_n = \langle \Phi_n^I  | \Psi \rangle .
  \label{prior_post_probls}
\end{eqnarray} 
Due to post-prior symmetry, the amplitudes in the two representations are related by
\begin{eqnarray}
  {\tilde c}^T_n =  c^T_n + \sum_m \langle \Phi_n^T  | \Phi_m^P \rangle c_m^P,  \nonumber\\ 
  {\tilde c}^P_n =  c^P_n + \sum_m \langle \Phi_n^P  | \Phi_m^T \rangle c_m^T.
  \label{prior_post_relations}
\end{eqnarray} 
The probabilities to populate the different final channels are defined as
\begin{equation}
  {\cal P}_j^{(T,P)} = |c_j^{(T,P)}|^2
  \label{CC_final_prob-prior}
\end{equation}
in the prior representation, or as
\begin{equation}
  {\tilde {\cal P}}_j^{(T,P)} = |{\tilde c}_j^{(T,P)}|^2 ,
  \label{CC_final_prob-post}
\end{equation}
in the post representation.
The index $j$ denotes a label to the quantum number of the final state in one of the two wells.

Because of the non-orthogonality of the basis states, the sum of these ``probabilities'' 
is not conserved during the collision.
If we instead define the ``probabilities'' by 
\begin{equation}
  {\cal P}_j^{(T,P)} = Re\left[ (c_j^{(T,P)})^*{\tilde c}_j^{(T,P)} \right],
  \label{CC_final_prob-unit}
\end{equation}
conservation of total probability is always fulfilled within the coupled-channel formalism.
This follows from the fact that the matrix governing the time evolution of the amplitudes in the post representation
is minus the Hermitian conjugate of the matrix that determines the time evolution of the amplitudes in the prior representation.
We shall therefore call equation (\ref{CC_final_prob-unit}) the \textit{unitary representation} of probabilities.
However,  there is  no guarantee that these quantities are always non-negative during the collision.

After the collision, when all overlaps between the basis states in the two wells vanish, 
the amplitudes for a given transition are the same in the post and prior representation,
as evident from equation (\ref{prior_post_relations}).
This so-called post-prior symmetry implies that the total  probability is conserved once the collision is over, 
also in a truncated coupled-channel treatment.
\subsubsection{Inclusion of continuum in the coupled-channels method \label{continuum}}
In reference \cite{Esbensen} only bound states were included in the bases and, hence, breakup channels were omitted. This is possibly justified for tightly bound systems, but not for weakly bound ones, 
for which the coupling to these channels can be very important.

In a coupled-channels scheme one can not include the full  continuum spectrum, since these states form a continuum of energies. Moreover, the fact that these states are not square-integrable poses numerical problems since the coupling potentials become of infinite range. To overcome these difficulties, it is customary to resort to an approximate, discrete description of the continuum. 
To describe the breakup process in CC a set of discretized continuum states defined on a finite spatial range is usually included for one or more subsystems of the reaction constituents. Even if the description of the process is supposed to occur at the end of the reaction, when the distance between the constituents tends to infinity, this constitutes a nonphysical situation. In fact, a proper description of the continuum, even if discretized, should take into account all the fragments involved in the reaction. This is what is done in the Faddeev formalism \cite{Faddeev}.

In time-dependent approaches, like the one we apply, to include a proper description of the continuum is a challenge. Here, we use the pseudostates introduced in \cite{MPV2016};
in particular, we discretize the continuum in an infinite square well basis (BOX method in \cite{MPV2016}).
We apply the same approach as in CDCC and provide discretized continuum pseudostates for each potential well. To include them in our coupled-channels calculation, 
we should ensure that at the end of the time evolution there is no overlap between target and projectile bases.
This is due to the fact that the two bases are not mutually orthogonal, 
so while they overlap the problem loses unitarity, as we mentioned in section \ref{CC}.
After the reaction, when the projectile is far enough for these overlaps to vanish, 
the problem has restored its unitarity and we can evaluate the final probabilities.
Thus, if we include the continuum we need to restrict it to a small range $[-x_b^J,+x_b^J]$ centered 
in the corresponding potential $J = \{\textit{target}, \textit{projectile}\}$, in order that $x_b^T + x_b^P >  x_P(t_{f})$.
A limitation of this method is that the choice of $x_b^J$ is done \textit{a posteriori} 
to include the breakup component into the continuum interval.
In situations in which the breakup channel is strongly dominant, it could become impossible to apply this method because the continuum component of the wavefunction might not be localized close to one of the potential wells. By comparing our CC results to the exact solution of the Schr\"odinger equation in this simple 1D framework, we are able to  understand the  limitation of this  description of the continuum.
\section{Numerical results: comparison of exact and approximate methods}\label{results}
In previous works we have applied the present approach on a variety of model cases 
to study different aspects of the direct reaction mechanism, 
always considering bound and weakly-bound systems. 
In particular, we have tested the model by varying the 
distance of closest approach between the two potential wells and the Q-value of the reaction in 
\cite{Moschini_thesis:2017,M2014,VM2015},
and we have studied the role of continuum in different configurations in \cite{Moschini_thesis:2017,MVM2016,MVA2018}. To explore different kinematical situations, we have performed further calculations.
In each case, we present the exact time-dependent solution (section \ref{exact}) and
 compare this result with the approximate solutions 
obtained with the coupled-channels calculation presented, as described in section  \ref{CC}.
The chosen model cases have been selected to illustrate several physical situations in which different reaction channels are favoured. For each case we have highlighted an analogous problem solved in a three dimensional CDCC environment, discussing the main challenges experienced in the 3D situation within our simplified framework.

\subsection{Case A \label{caseA}}
\begin{figure}[!hbt]
\begin{center}
\includegraphics[scale=.45]{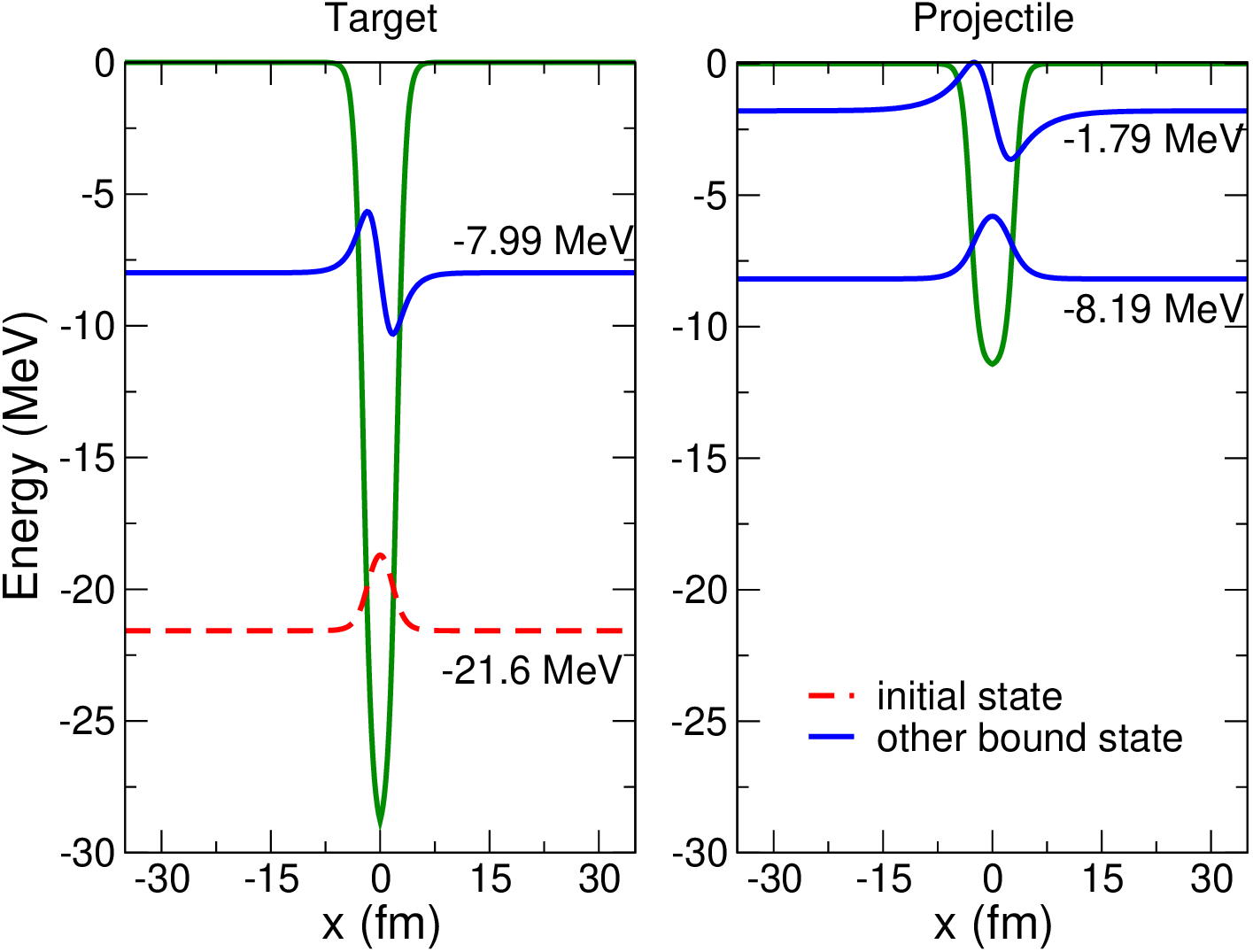}
\caption{Target (left) and projectile (right) Woods-Saxon potential and corresponding bound states 
  for the case A. The initial state, in this case the target ground state, is the dashed red curve.}
\label{fig_a_potentials}
\end{center}
\end{figure}

We start with the simplest case of a well bound target in which we observe a dominance of elastic and inelastic channels.
The target and projectile potentials are depicted in figure \ref{fig_a_potentials} 
with their respective bound states wavefunctions.
The lowest state in the target is chosen as initial state for this reaction, its wave function is shown as dashed red curve in figure \ref{fig_a_potentials}.
The projectile follows the trajectory (\ref{esbensen_trajectory})
with an asymptotic velocity of $0.1 \times c$, which corresponds to an asymptotic energy of $5.0$ MeV.
The reduced mass for this systems is $1.001$ amu.

\begin{figure}[!hbt]
\centering
\subfigure[]{\includegraphics[scale=0.24]{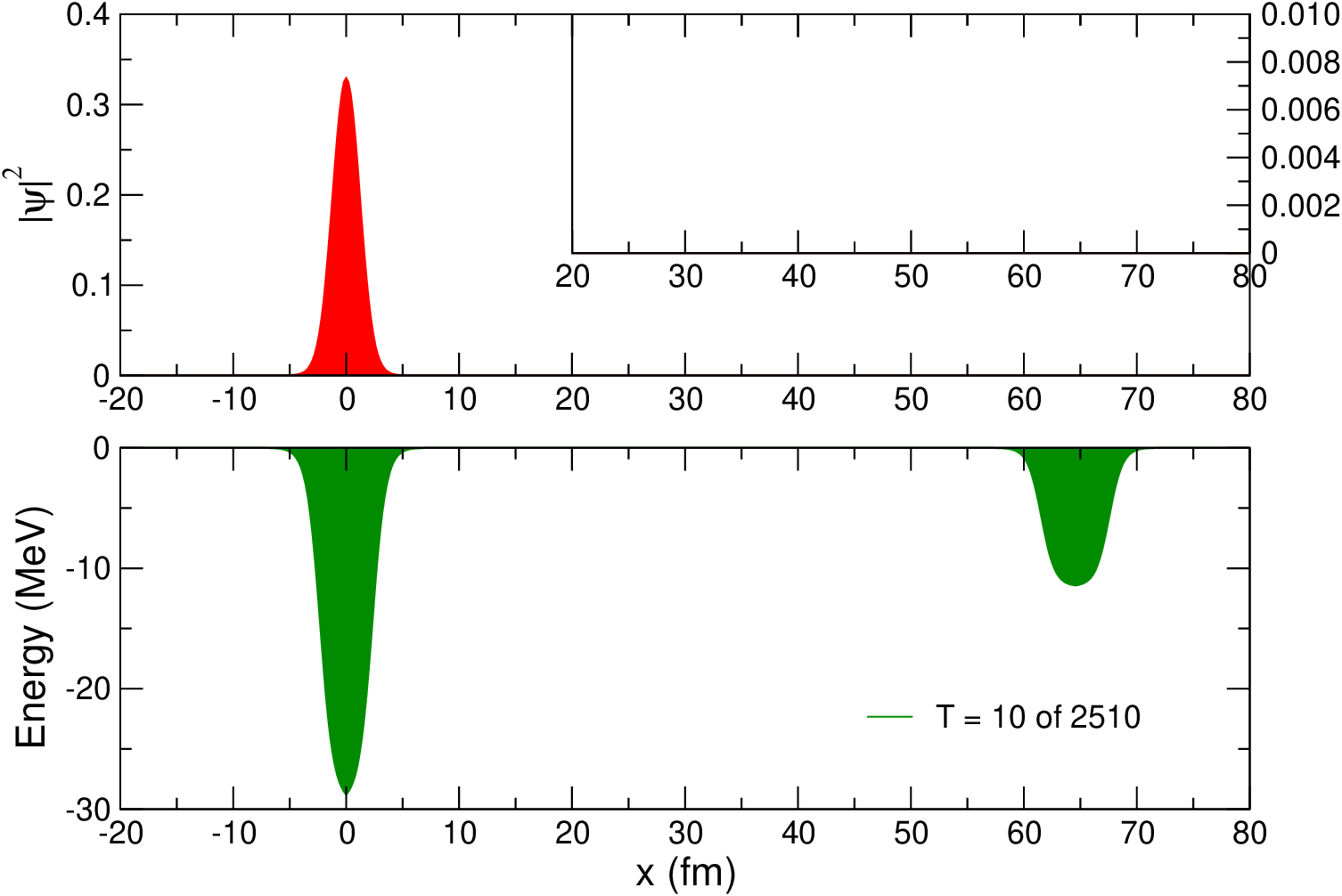}}
\subfigure[]{\includegraphics[scale=0.24]{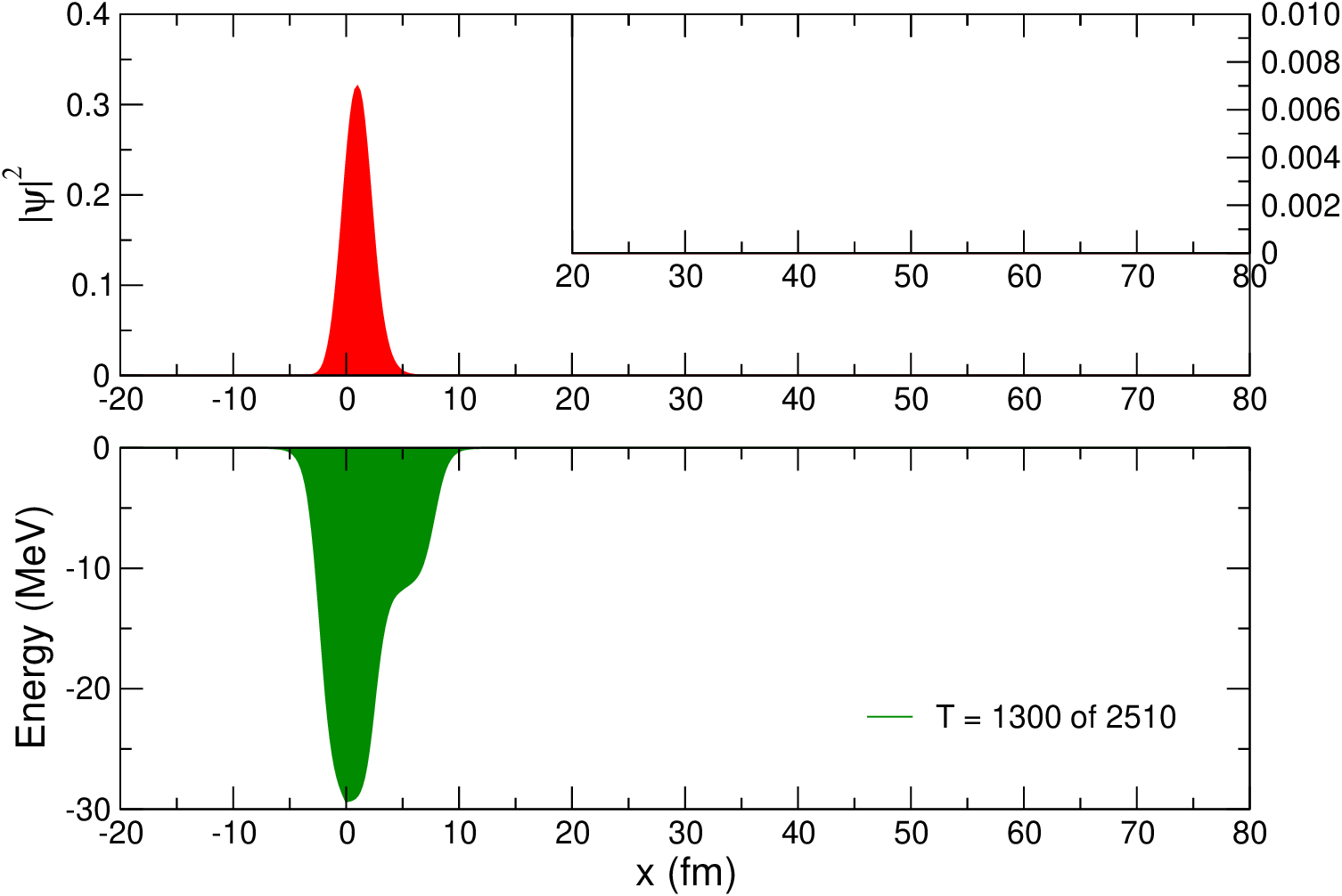}}
\vspace{0.1cm}
\subfigure[]{\includegraphics[scale=0.24]{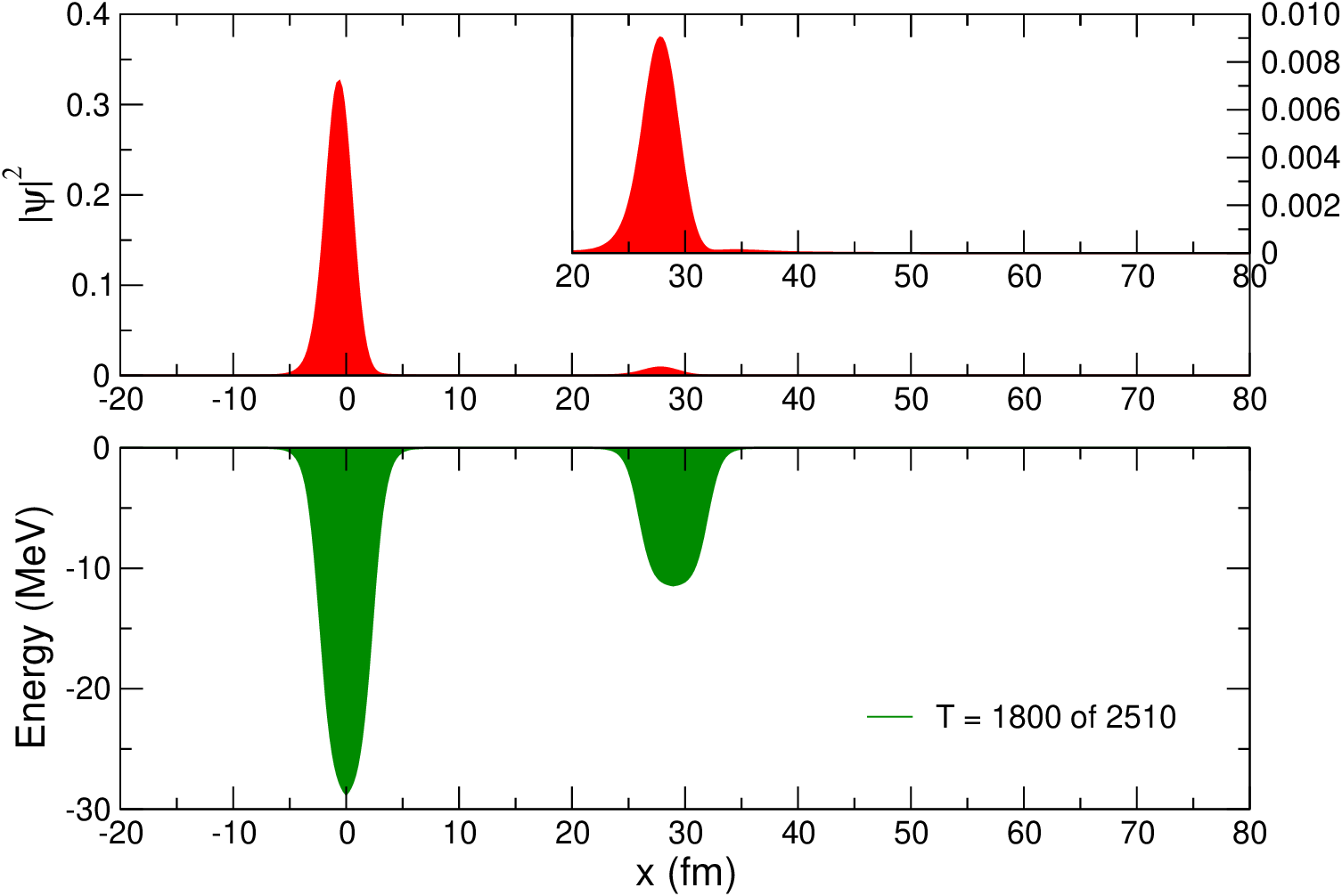}}
\subfigure[]{\includegraphics[scale=0.24]{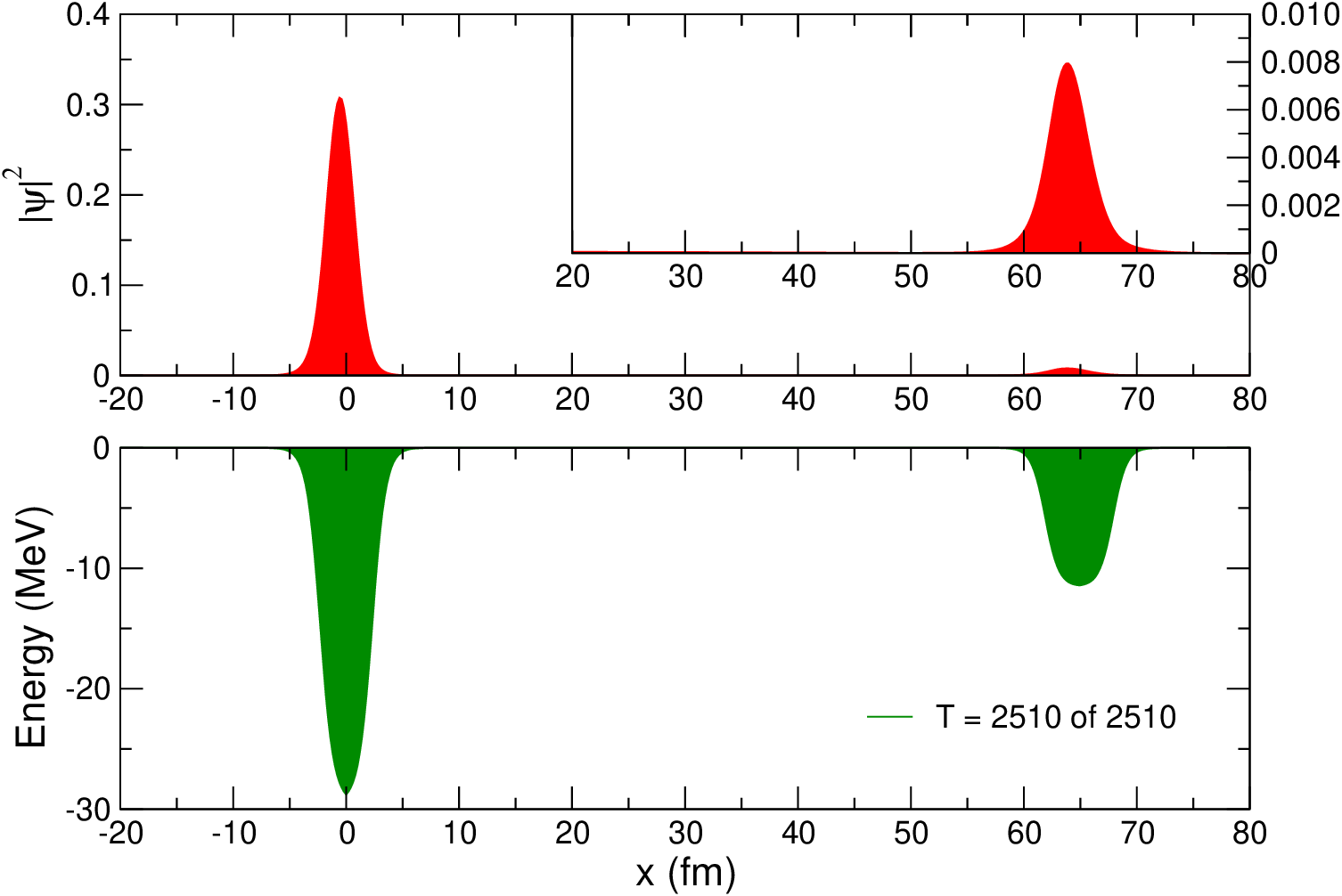}}
\caption{\label{f1} Model case A.
      Upper panels: the squared total wavefunction at different times with exact method 
      and a zoom for $x>20$~fm in the inset.
      Lower panels: the target and projectile potential wells. 
      The elapsed time for each frame is indicated in the legends. }
\end{figure}

In figure \ref{f1} we present the potentials 
and the exact wavefunction squared at different moments of the time evolution.
In each frame the upper panel shows the exact wavefunction 
and the lower displays the target and projectiles potentials.
One can see that a small component of the system wavefunction is transferred 
and remains bound to the projectile potential well as it moves away after the collision.

In table \ref{tab2} we present the results obtained by solving 
the problem using different approaches and different bases. 
\begin{table}[h!]
\centering
\begin{tabular}{c|c|c|c|c|c}
\hline
              & Exact  & CC$^{(1)}$ & CC$^{(2)}$ & CC$^{(3)}$ & CC$^{(4)}$\\  
\hline
Elastic       & 86.1\% & 89.8\%    &    89.4\% & 74.1\%   & 84\%      \\
Inelastic     & 9.8\%  & 10.1\%    &    10.3\% & 18.4\%   & 11.6\%    \\ 
\hline
Transfer g.s. & 2.5\%  & -         &    -      & 6.8\%    & 2.3\%     \\ 
Transfer 1    & 0.8\%  & -         &    -      & 0.56\%   & 1.9\%     \\
\hline
Breakup       & 0.7\%  & -         &    0.3\%  & -        &  0.13\%    \\
\hline
\end{tabular}
\caption{Final probabilities for the model case A
    obtained with the exact and the coupled-channels methods.}
\label{tab2}
\end{table}
In particular, they correspond to:
    \begin{itemize}
        \item \textbf{Exact:} Calculation done by solving numerically the time-dependent Schr\"odinger equation 
                             (\ref{Schr\"odinger_equation_time_dependent}).
                             The different results represent the probability for the valence neutron to remain in the ground state 
                                    (Elastic) or to directly excite the corresponding bound states of the two potential wells
                                    (``Inelastic'' and ``Transfer'') or to excite continuum states (Breakup).
        \item \textbf{CC$^{(1)}$:} Coupled-channels calculation in which only the two target bound states are included.
        \item \textbf{CC$^{(2)}$:} Coupled-channels calculation using target bound levels 
                                 plus the first 10 continuum pseudostates obtained 
                                in a range equal to the maximum radius of the grid.
        \item \textbf{CC$^{(3)}$:} Coupled-channels calculation including only target and projectile bound states.
        \item \textbf{CC$^{(4)}$:} Coupled-channels calculation including target and projectile bases composed 
                                 by bound and 5 continuum pseudostates 
                                 calculated in a $\left[-20;20\right]$ fm range centered in 
                                 the respective potential. 
                                 It corresponds to an energy cutoff in the continuum of $4$ MeV.   
    \end{itemize}
By comparing the results of table \ref{tab2}, we note that 
the coupled-channels calculation without transfer and continuum channels, 
i.e.\ \textbf{CC$^{(1)}$}, reproduces rather well the elastic and inelastic probabilities. 
This is a consequence of the dominance of the elastic channel and, to a lesser extent, the inelastic channel in this case.
Including the breakup (\textbf{CC$^{(2)}$}) channel alone does not improve the result, because it is not possible to describe the transfer in this configuration.
Interestingly, a calculation with elastic, inelastic, and transfer channels  (\textbf{CC$^{(3)}$}) 
is not enough to reproduce the expected values, even if those are the dominant channels in this case.
By including all possible channels simultaneously (\textbf{CC$^{(4)}$}) we obtain a fine agreement with the numerical solution.
\begin{figure}[!hbt]
\begin{center}
\includegraphics[scale=.45]{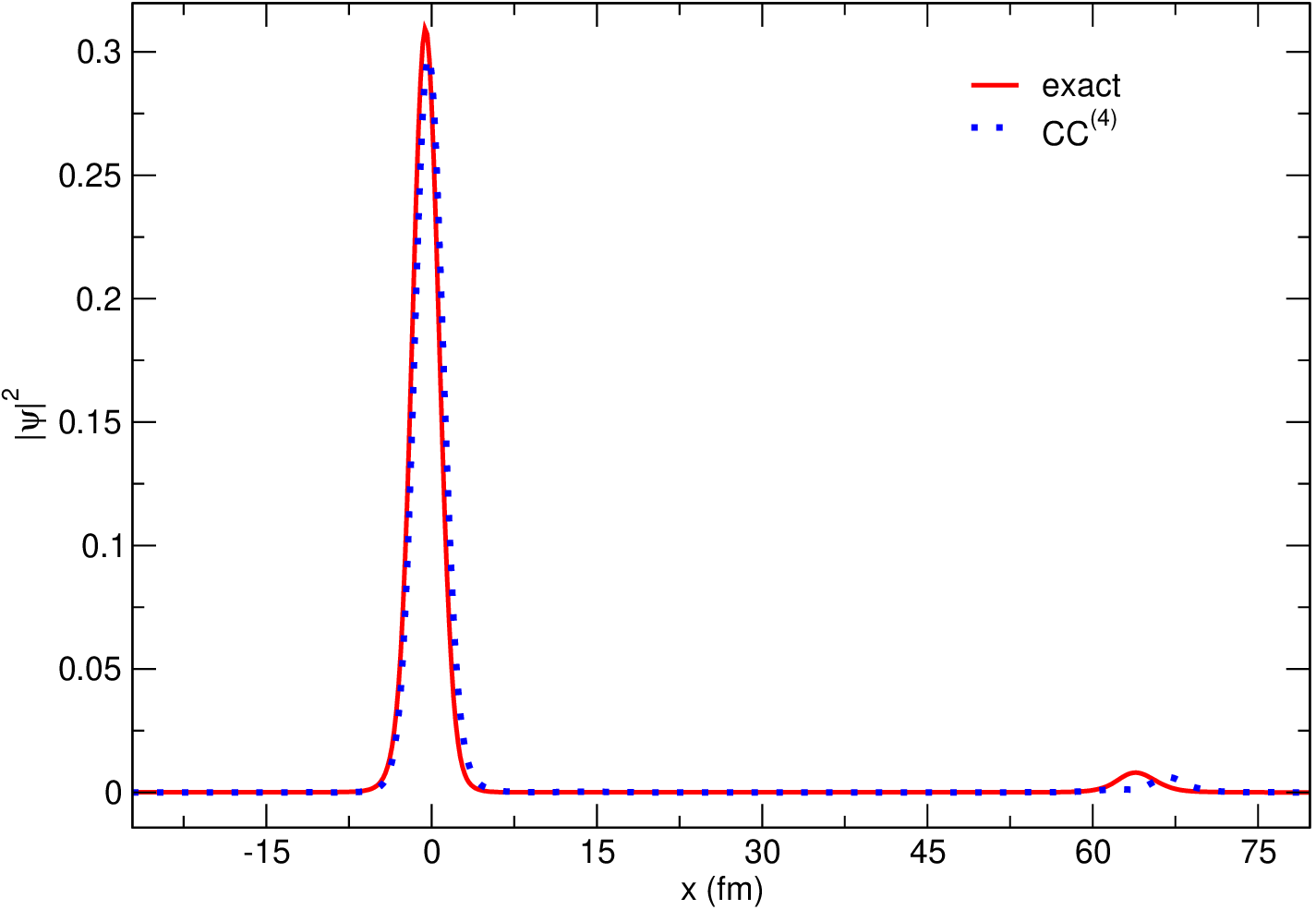}
\caption{Final ``exact'' wavefunction (solid red line) and final \textbf{CC$^{(4)}$} wavefunction  (dotted blue line). The projectile potential is placed at $x_P(t_f)=65$~fm.}
\label{figWF-A}
\end{center}
\end{figure}

\begin{figure}[!hbt]
\begin{center}
\includegraphics[scale=.45]{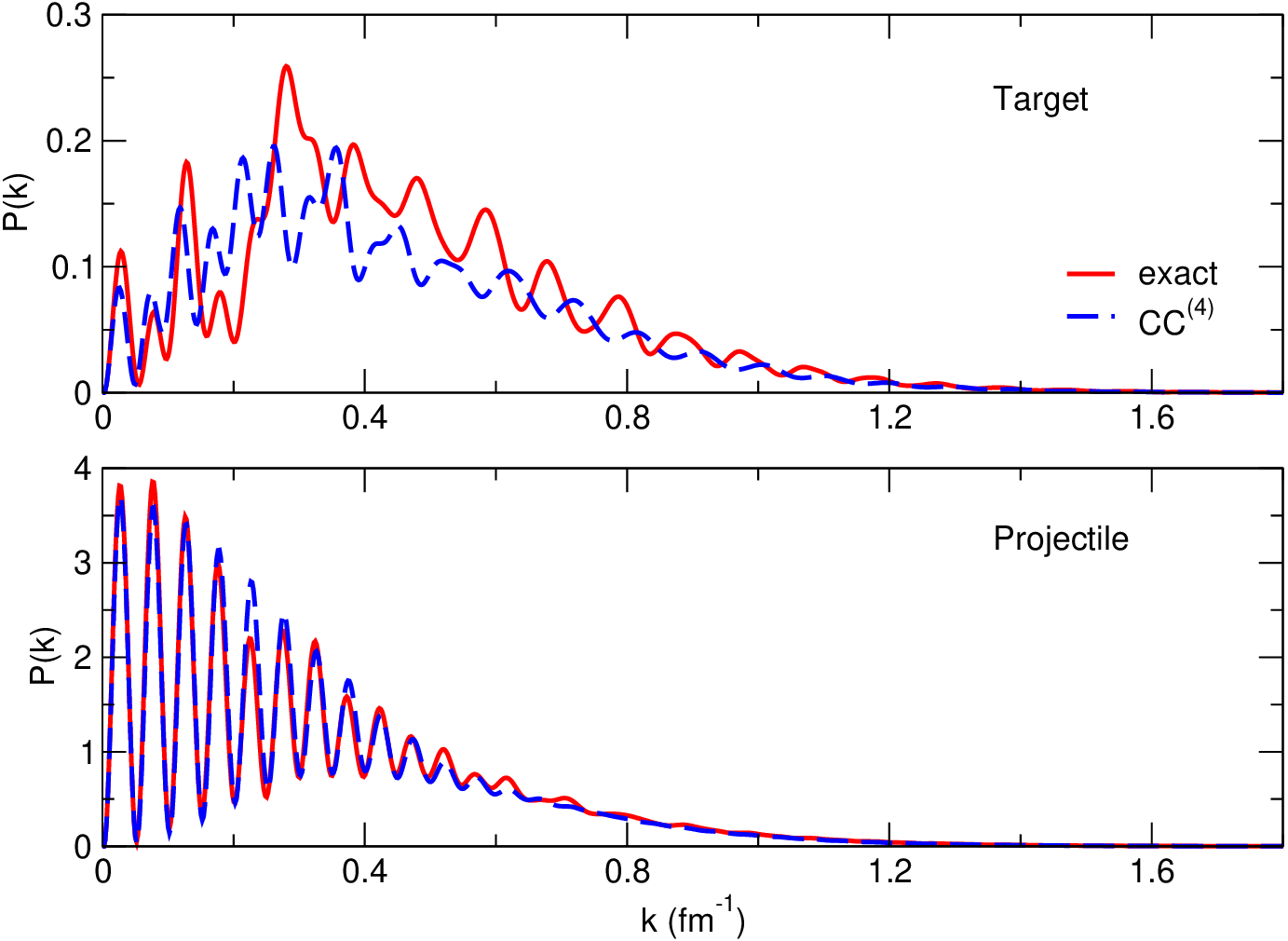}
\caption{Breakup probabilities as a function of the momentum in the continuum. The upper and lower panels correspond to the projection on target and projectile states, respectively. Solid red lines indicate the ``exact'' final wavefunction, while dashed blue lines indicate the \textbf{CC$^{(4)}$} final one.}
\label{fig-excont-A}
\end{center}
\end{figure}

In figure \ref{figWF-A}, we compare the final ``exact'' wavefunction (solid red line), already reported in the upper panel of figure \ref{f1}(d), to the wavefunction obtained from calculation \textbf{CC$^{(4)}$} using  the final coupled-channels coefficients in eq.~(\ref{wf_CC_combination}) (dotted blue line). We observe quite good an agreement between the two results, the slight differences being due to the overestimation of the excitation of the first excited states of both target and projectile wells by the CC calculation. 
To further understand the role of continuum states in this process, we construct the ``exact'' momentum-normalized bases $\varphi(k,x)$ for the two potential wells. We then calculate the probability as a function of the momentum ${\cal P}(k)$ following eq.~(\ref{breakup_continuum_overlap}). The results are displayed in figure \ref{fig-excont-A}: the exact (solid red) and  \textbf{CC$^{(4)}$} (dashed blue) final wavefunctions from figure \ref{figWF-A} have been projected onto the target and projectile exact continuum in the upper and lower panels, respectively. We can notice how the ``projectile continuum'' plays a more relevant role than the target continuum, and also how well the \textbf{CC$^{(4)}$} calculation works in this case. For the ``target continuum'' we observe that the coupled-channels calculation does not reproduce accurately the structure observed at lower momenta. Anyway, as already observed, the total strength is negligible with respect to the role of the other well. It should be noticed that the incident energy of $5$~MeV corresponds to a threshold of $k=0.49$~fm$^{-1}$; the excitation slightly exceeds this limit probably due to the non conservation of energy in semiclassical descriptions \cite{FB12}.
 
This case is ideally described within the CC method, given the dominance of the elastic channel. Many authors have found an excellent agreement with elastic and transfer experimental cross sections, provided that breakup channels are properly included in the CDCC calculation (see e.g.~\cite{YOMM12,MCNT02,LCAGQG09,KAKR09}).
Also, some authors have suggested that continuum-continuum couplings have a strong influence in the complete and incomplete fusion reaction for weakly-bound nuclei \cite{HVDL00,DT02,CLGH09}. 
The present model case confirms the crucial role of continuum, and contributes to demonstrate that even if the system under study is well bound and the breakup component is negligible, 
the inclusion of continuum states can be essential to get a proper description of the reaction.
In this situation in which the breakup component is not dominant, to include two sets of continuum states, each one related to one potential well, allows us to obtain a result in agreement with the expected exact one. So, this treatment of continuum could constitute an acceptable approximation in these conditions.

\subsection{Case B \label{caseB}}
\begin{figure}[!hbt]
\begin{center}
\includegraphics[scale=.35]{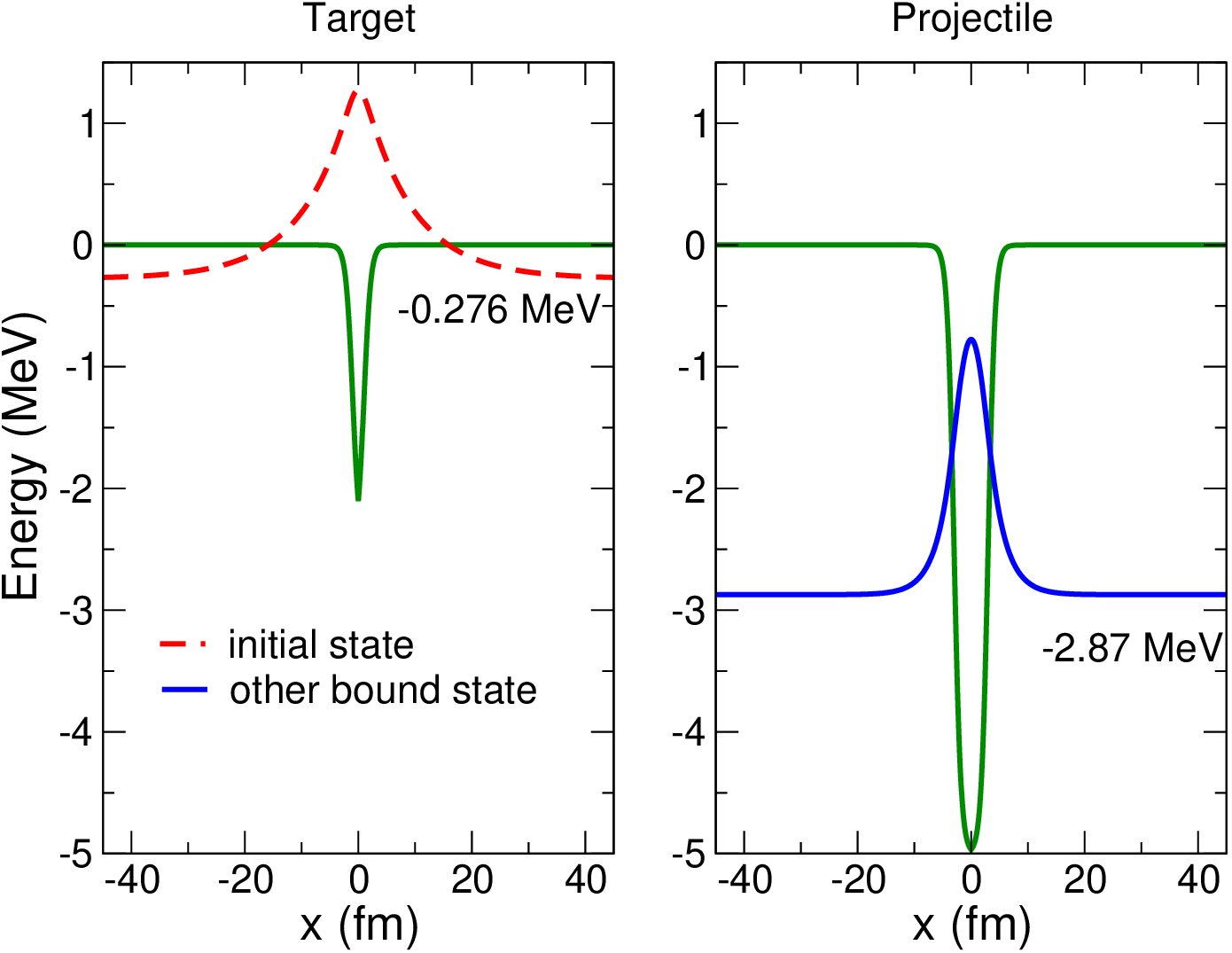}
\caption{Target (left) and projectile (right) Woods-Saxon potential and corresponding bound states for the model case B.
The initial state, in this case the target ground state, is the dashed red curve.}
\label{pot:caseB}
\end{center}
\end{figure}

In this second case the initial target state is extremely weakly-bound
and consequently the breakup channel is the most relevant.
The target and projectile potentials are depicted in figure \ref{pot:caseB} 
with their respective bound states wavefunctions. 
The initial state is the dashed red curve corresponding to the target bound state.
The energy of this state is $-2.276$~MeV and its weakly-bound nature is clearly evident from the extended
tail of its wavefunction.
The projectile follows the trajectory (\ref{esbensen_trajectory})
with asymptotic velocity of $0.1 \times c$ corresponding to an incident energy of $5.0$ MeV, 
and with a reduced mass of $1.001$ amu.

\begin{figure}[!hbt]
\centering
\subfigure[]{\includegraphics[scale=0.24]{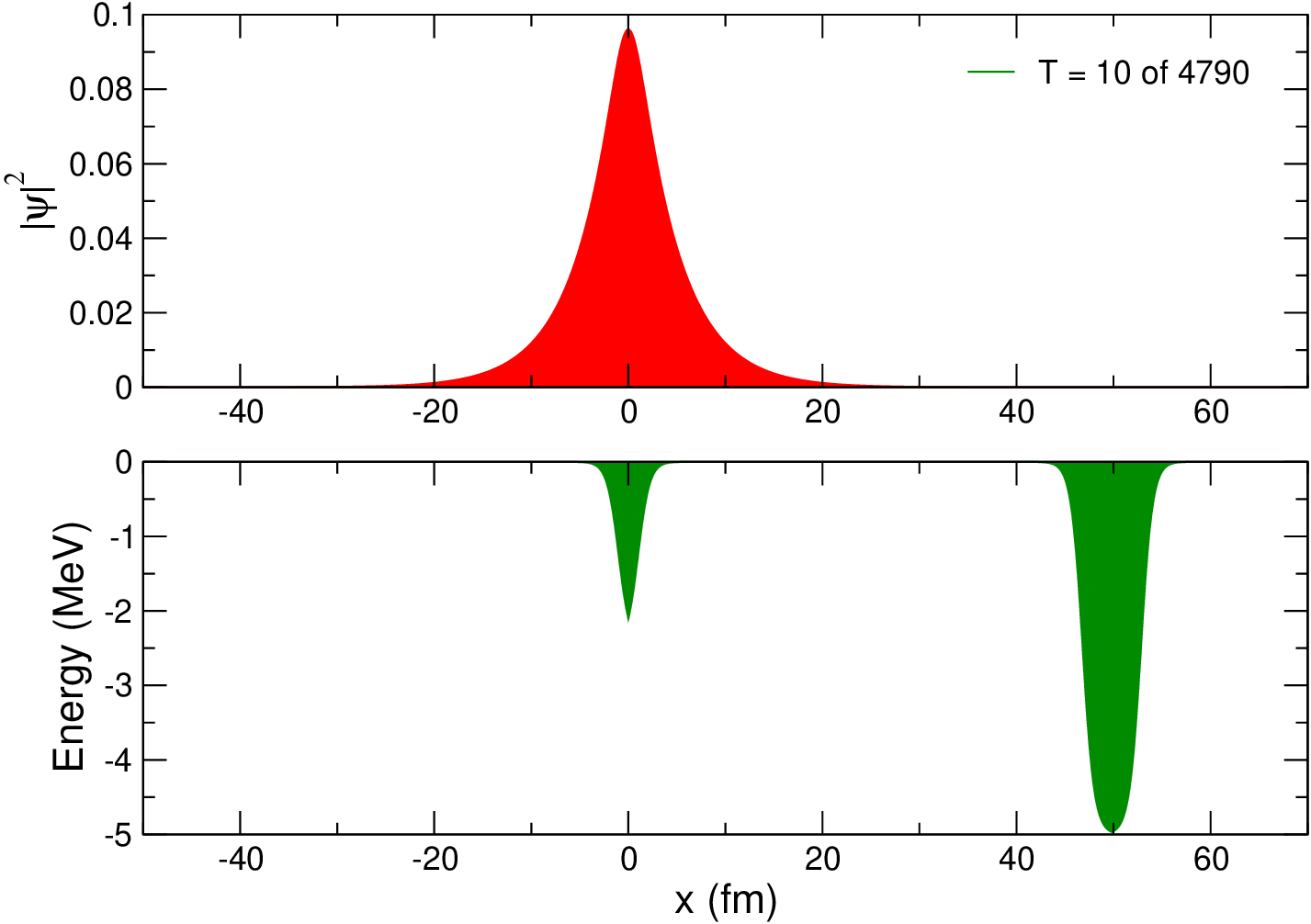}}
\subfigure[]{\includegraphics[scale=0.24]{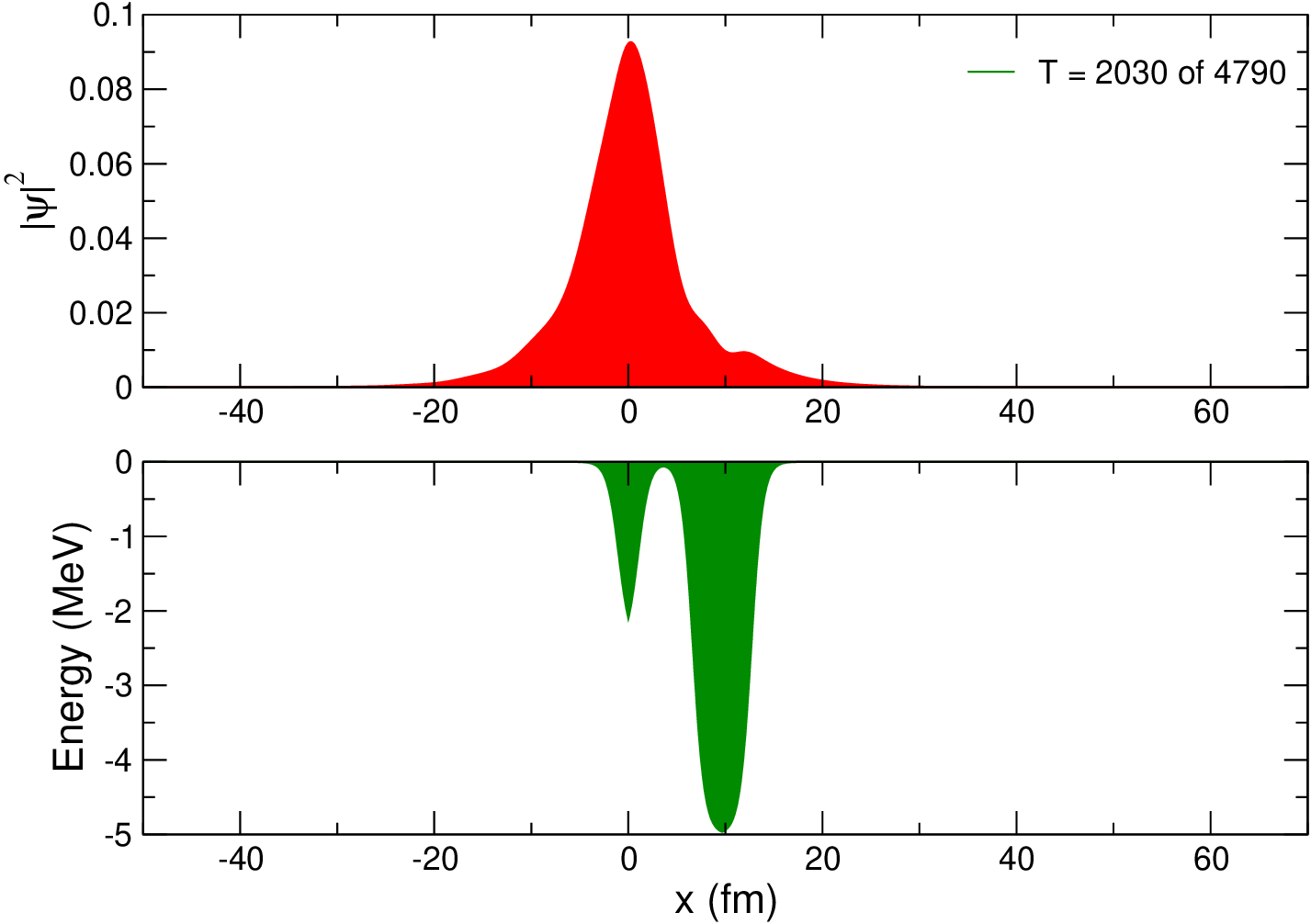}}
\subfigure[]{\includegraphics[scale=0.24]{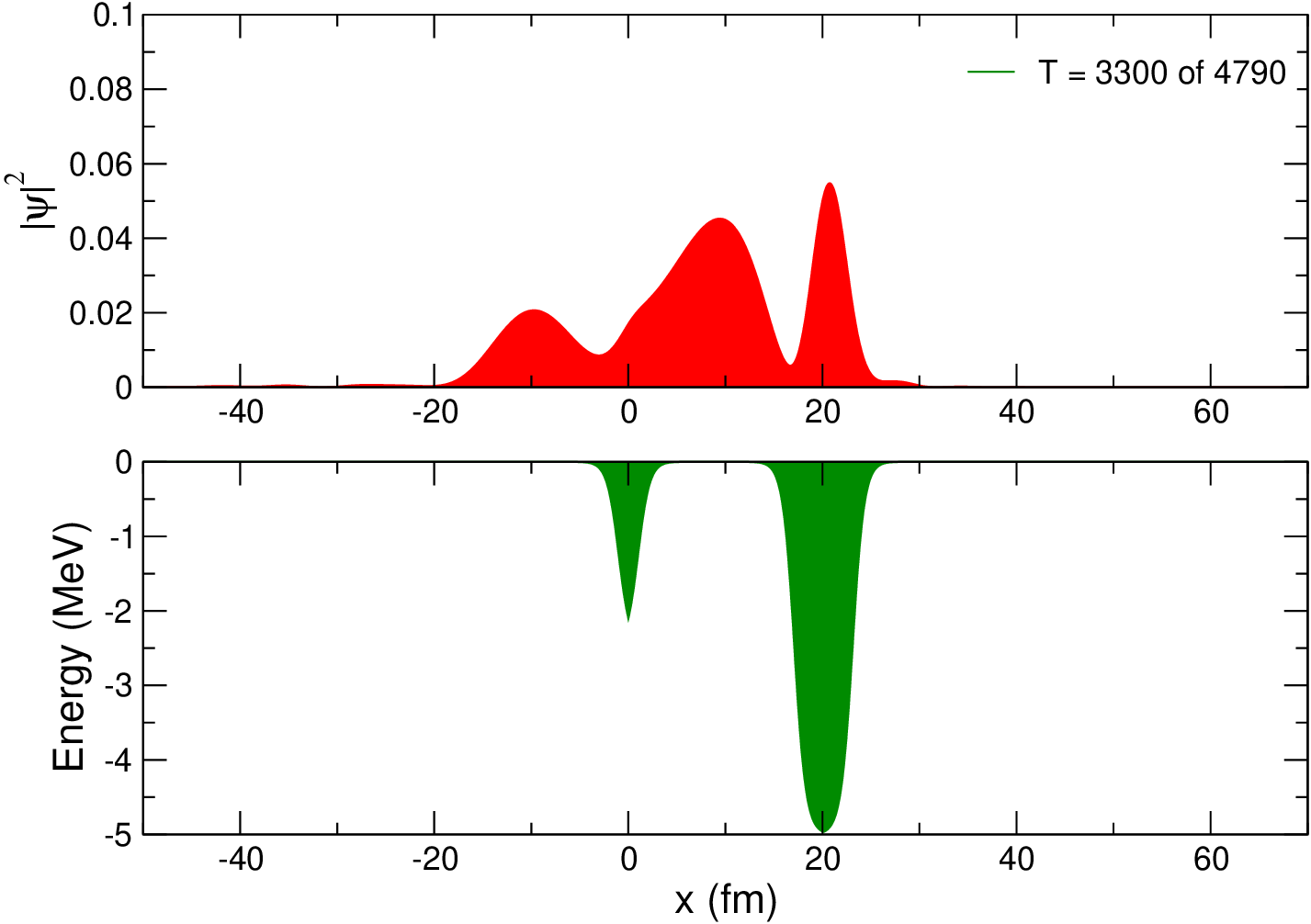}}
\subfigure[]{\includegraphics[scale=0.24]{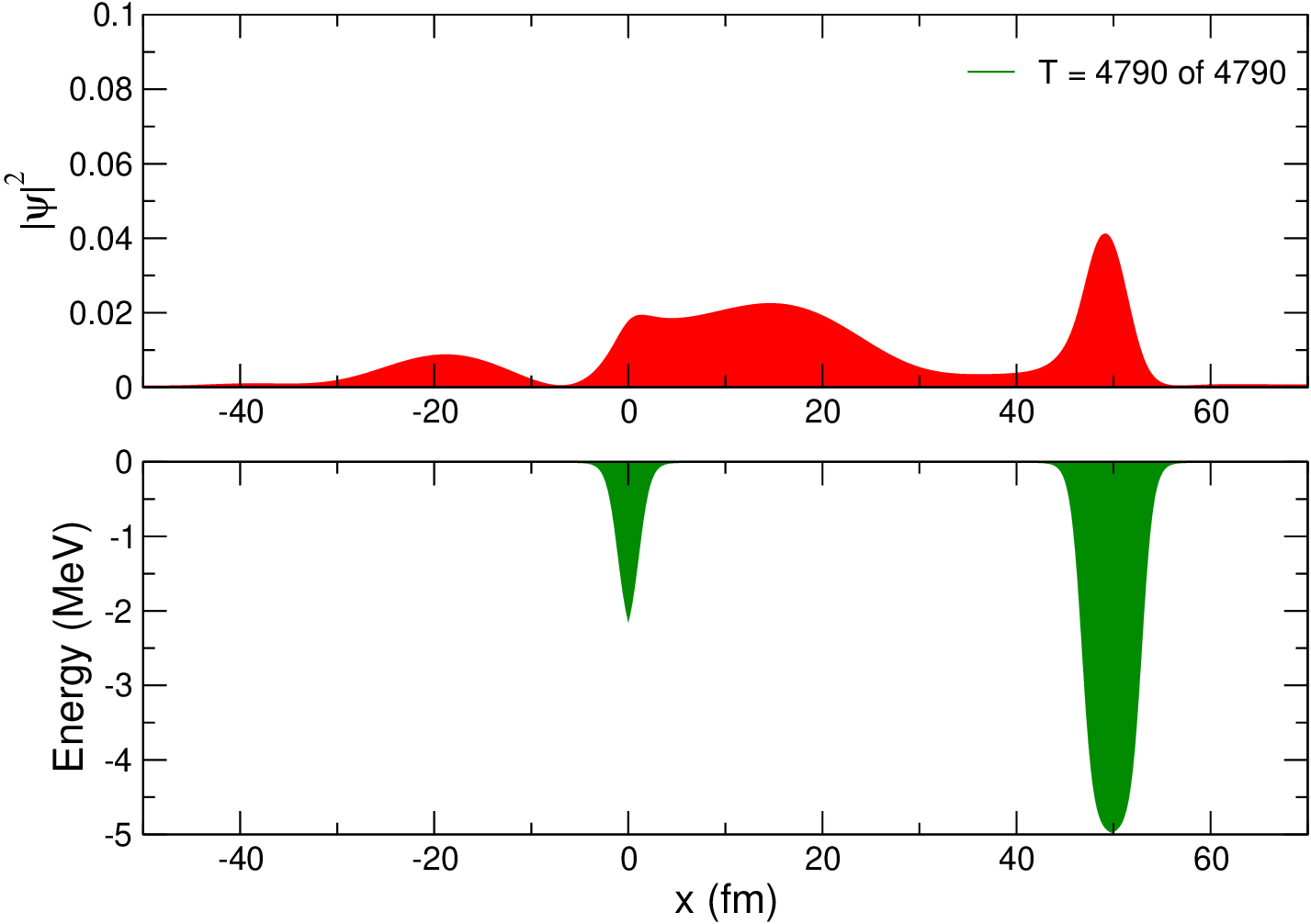}}
\caption{\label{f3} Model case B. 
      Upper panels: the squared total wavefunction at different times with exact method.
      Lower panels: the target and projectile potential wells. 
      The elapsed time for each frame is indicated in
the legends.
}
\end{figure}

The evolution of the exact wavefunction during the collision is presented in figure~\ref{f3},
where in each frame the upper panel shows the exact wavefunction 
and the lower one displays the target and projectiles potentials.
In this case, both the transfer and the continuum components of the system wavefunction are clearly evident.

The results for case B, obtained applying different methods, are presented in table~\ref{tab4}.   
\begin{table}[h!]
\centering
\begin{tabular}{c|c|c|c|c|c}

         & Exact  & CC$^{(1)}$ & CC$^{(2)}$   & CC$^{(3)}$\\
\hline
Elastic  & 20.94\%&    22.9\% &21.9\%       & 25.8\%\\
\hline
Transfer & 6.73\% &    -      & $<10^{-3}$\% &  0.3\%\\
\hline
Breakup  & 72.32\%&    77.1\% & 78.2\%      &  73\%\\
\hline
\end{tabular}
\caption{Final probabilities for the model case B
    obtained with the exact and the coupled-channels methods.}
\label{tab4}
\end{table}
In particular, they correspond to:
    \begin{itemize}
        \item \textbf{Exact:} Calculation done by solving numerically the time-dependent Schr\"odinger equation 
                             (\ref{Schr\"odinger_equation_time_dependent}).
                             The different results represent the probability for the valence neutron to remain in the ground state 
                                    (Elastic) or to excite the bound state of the projectile wells
                                    (Transfer) or to excite continuum states (Breakup).
        \item \textbf{CC$^{(1)}$:} Coupled-channels calculation in which the target bound state 
                                 and 10 target continuum states within the $\left[-40;40\right]$~fm range 
                                     are included.
        \item \textbf{CC$^{(2)}$:} Coupled-channels calculation using target and projectile bound levels 
                                 plus 50 continuum target pseudostates obtained in a radius equal to $40$~fm.
        \item \textbf{CC$^{(3)}$:} Coupled-channels calculation including target and projectile bases composed 
                                 by bound and 5 continuum pseudostates 
                                 calculated in a $\left[-22;22\right]$~fm range centered in 
                                 the respective potential. 
                                 It corresponds to an energy cutoff in the continuum of about $4$ MeV.   
    \end{itemize}
From the calculation \textbf{CC$^{(1)}$} it is evident that the target basis  
can not describe alone the ``exact'' result, even if the proportion between exact and breakup channels is 
reproduced quite well.
By adding the projectile bound state (\textbf{CC$^{(2)}$}) the elastic channel is better reproduced, 
but the transfer is not well accounted for. Including also the projectile continuum improves the result (\textbf{CC$^{(3)}$}), 
thus showing that in a perturbation picture the transfer is not a one-step process, 
but it is reached through successive steps involving the continuum.
Multi-step processes in direct reactions are traditionally inferred by CDCC calculations, 
and, as we have seen in the previous section, continuum--continuum couplings play a crucial role \cite{DT02,NT99,LM19}.
\begin{figure}[!hbt]
\begin{center}
\includegraphics[scale=.45]{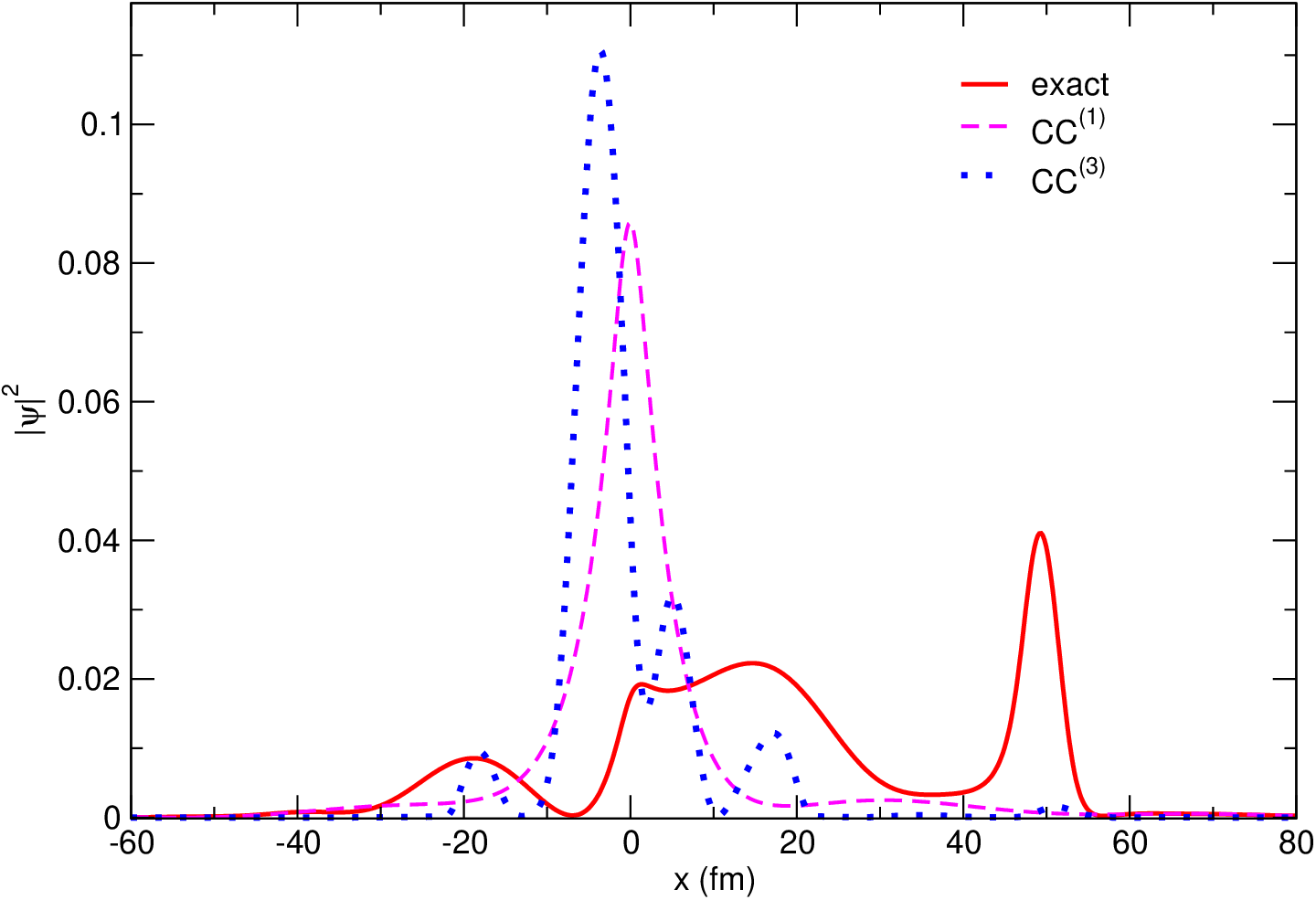}
\caption{Final ``exact'' wavefunction (solid red line), final \textbf{CC$^{(1)}$}  (dashed magenta line) and final \textbf{CC$^{(3)}$}   (dotted blue line) wavefunctions. The projectile potential is placed at $x_P(t_f)=50$~fm.}
\label{figWF-B}
\end{center}
\end{figure}
\begin{figure}[!hbt]
\begin{center}
\includegraphics[scale=.45]{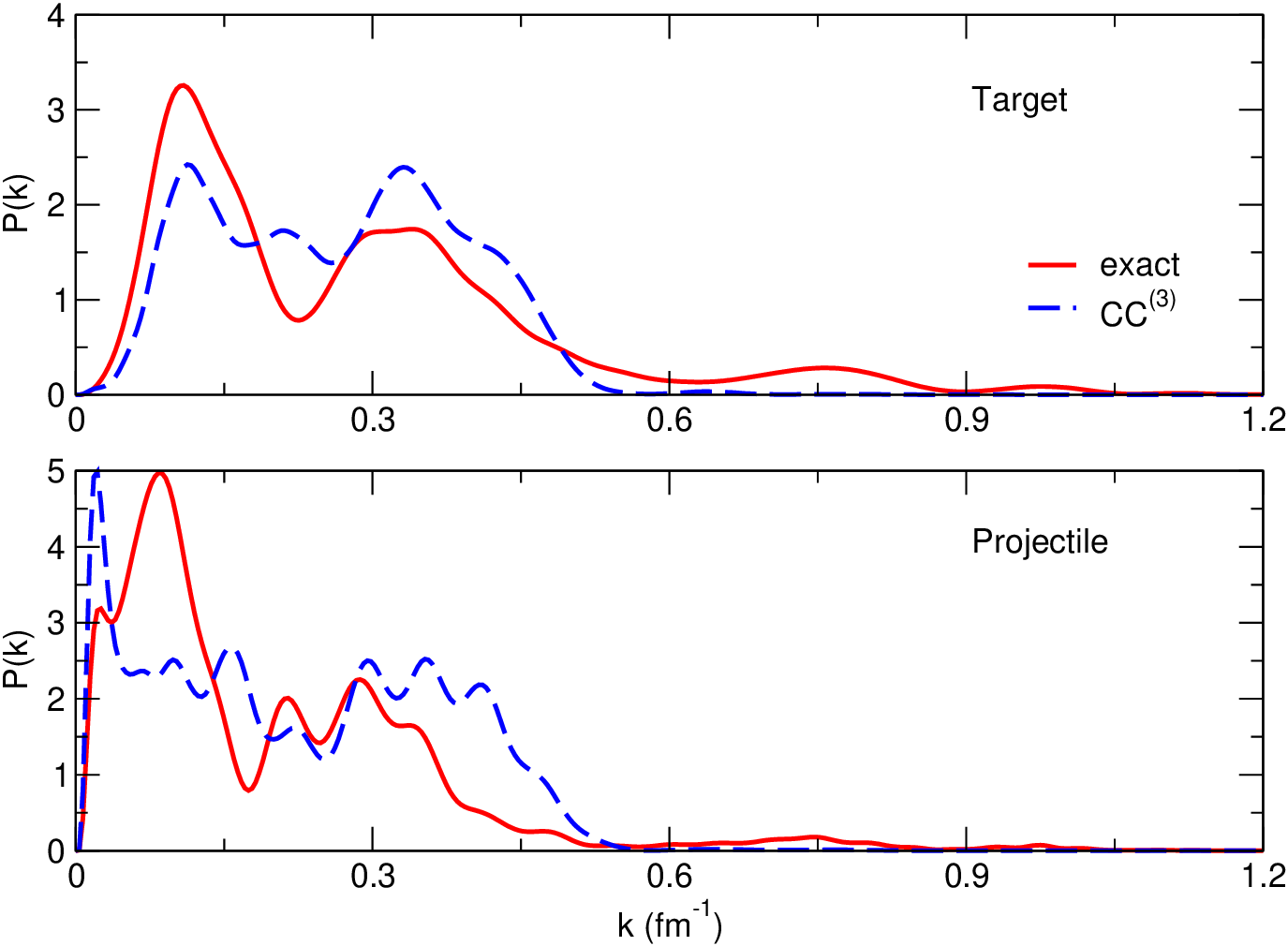}
\caption{Breakup probabilities as a function of the momentum in the continuum. The upper and lower panels correspond to the projection on target and projectile states, respectively. Solid red lines indicate the ``exact'' final wavefunction, while dashed blue lines indicate the \textbf{CC$^{(3)}$} final one.}
\label{fig-excont-B}
\end{center}
\end{figure}

In figure \ref{figWF-B} we show the final wavefunctions for the exact (solid red), \textbf{CC$^{(1)}$} (dashed magenta), and \textbf{CC$^{(3)}$}   (dotted blue) calculations. Neither of the two CC wavefunctions reproduces the shape of the final exact wavefunction, although their breakup probabilities are in agreement. 
This is partly explained by the failure in describing the transfer component, and, more importantly, by the spatial restriction of the pseudostates.
 This is also seen in figure \ref{fig-excont-B}, where we show the projection of the exact (solid red) and \textbf{CC$^{(3)}$} (dashed blue) wavefunctions onto the exact continuum $\varphi(k,x)$ of target and projectile wells (upper and lower panels, respectively). The coupled-channels calculation does not reproduce properly the exact result: only their  magnitudes are in agreement and, in the case of the ``target continuum'', the main structure is reproduced with the maximum and minimum values placed at similar momenta. Anyway, the similar order of magnitude of the target and projectile distributions suggests the importance of including the continuum of all the possible subsystems of the outgoing particles to properly describe all the reaction channels. Moreover they emphasize the fact that, when both the projectile and target systems are weakly bound, the choice for the continuum representation is not obvious and might strongly affect the final result. We notice also that, since both basis are not mutually orthogonal, including both representations leads to  overcompleteness, similar to what is done in the Faddeev formalism \cite{Faddeev}.

To further improve the coupled-channels solution we should add more continuum pseudostates 
and enlarge the range over which they are defined. 
Doing so, the two bases would overlap and we would obtain a non-unitarity solution (as we detailed in the previous section).
To avoid that, we should let the projectile evolve further, 
but in this way the continuum component of the system wavefunction would 
get away from the two wells and none of the two continuum representations would describe properly the process.
As one can notice, already in the \textbf{CC$^{(3)}$} calculation a small component of the system wavefunction has not been included in the 
range covered by the two bases (around $-30$~fm in Fig.~\ref{f3}d).
Anyway, even if the ``exact'' result is not fully reproduced, 
the coupled-channels calculation allows us to infer important information about the reaction mechanism 
for this case, and in particular about the transfer process.
This resembles the case involving weakly bound systems, in which  breakup becomes an important reaction channel and the CDCC description  becomes challenging.  Also in three dimensional case, the discrete basis representing the projectile and/or target continuum is spatially constrained by the extension of the basis, and hence  the calculated breakup observables are expected to be reliable only in those kinematical situations which are sensitive to that region \cite{DMCNF}.

\subsection{Case C}\label{caseC}
We treat now a case in which all the possible channels are relevant: 
elastic, inelastic, transfer and breakup.
The target and projectile potentials are depicted in figure \ref{pot:caseC} 
with their respective bound-state wavefunctions. 
The initial state is represented by the red  dashed curve
corresponding to the first excited state of the target. We assume that the target ground-state is Pauli forbidden 
and the corresponding channel should not be included in the calculation.
However, in this case we take into account this channel, calling it ``decay to g.s.''.
The projectile follows the trajectory (\ref{esbensen_trajectory})
with asymptotic velocity of $0.05 \times c$ corresponding to an incident energy of $1.0$ MeV, 
and with a reduced mass of $0.975$ amu.

\begin{figure}[!htbp]
  \centering
    \subfigure[]{\includegraphics[scale=.21]{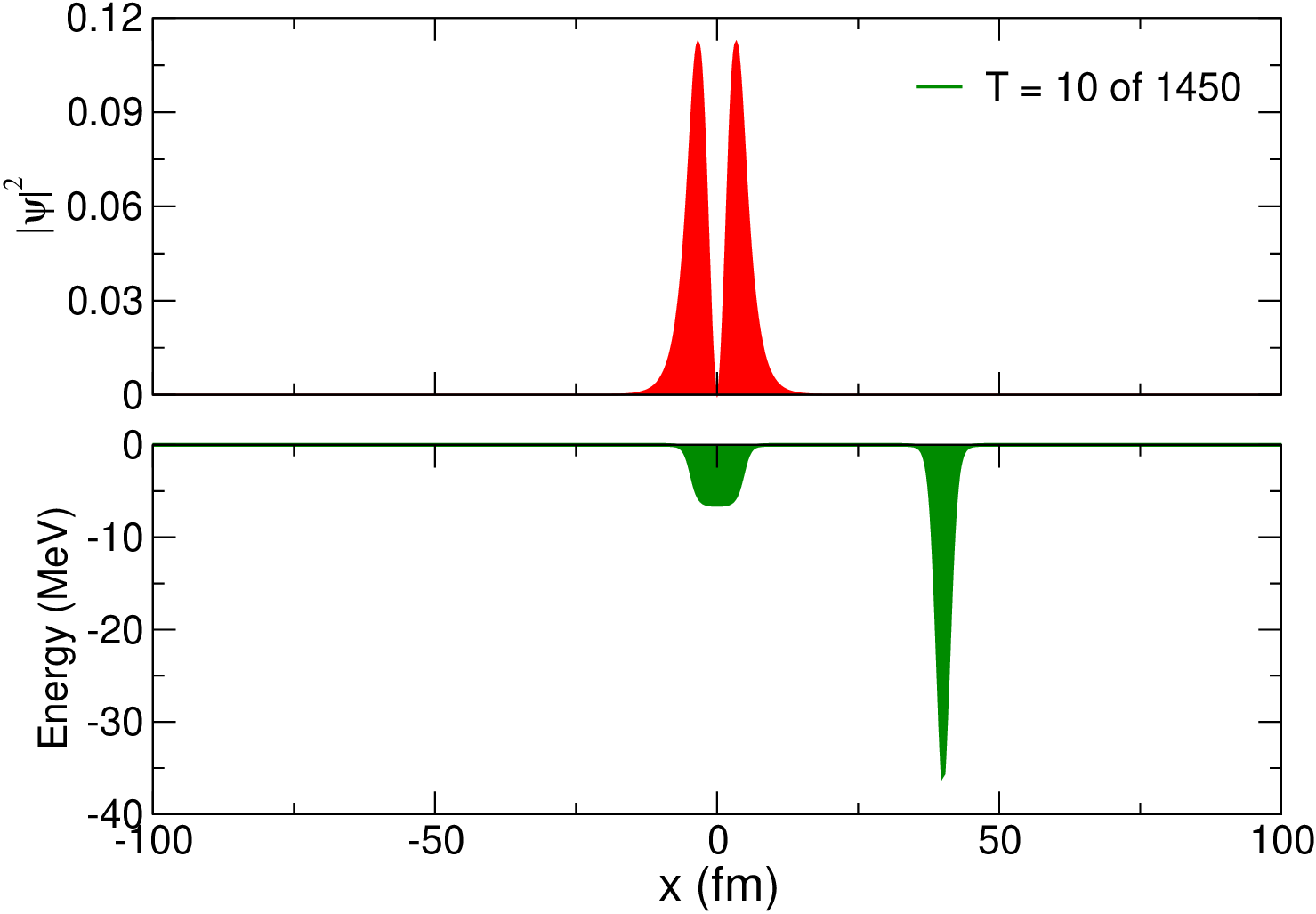}}
    \subfigure[]{\includegraphics[scale=.21]{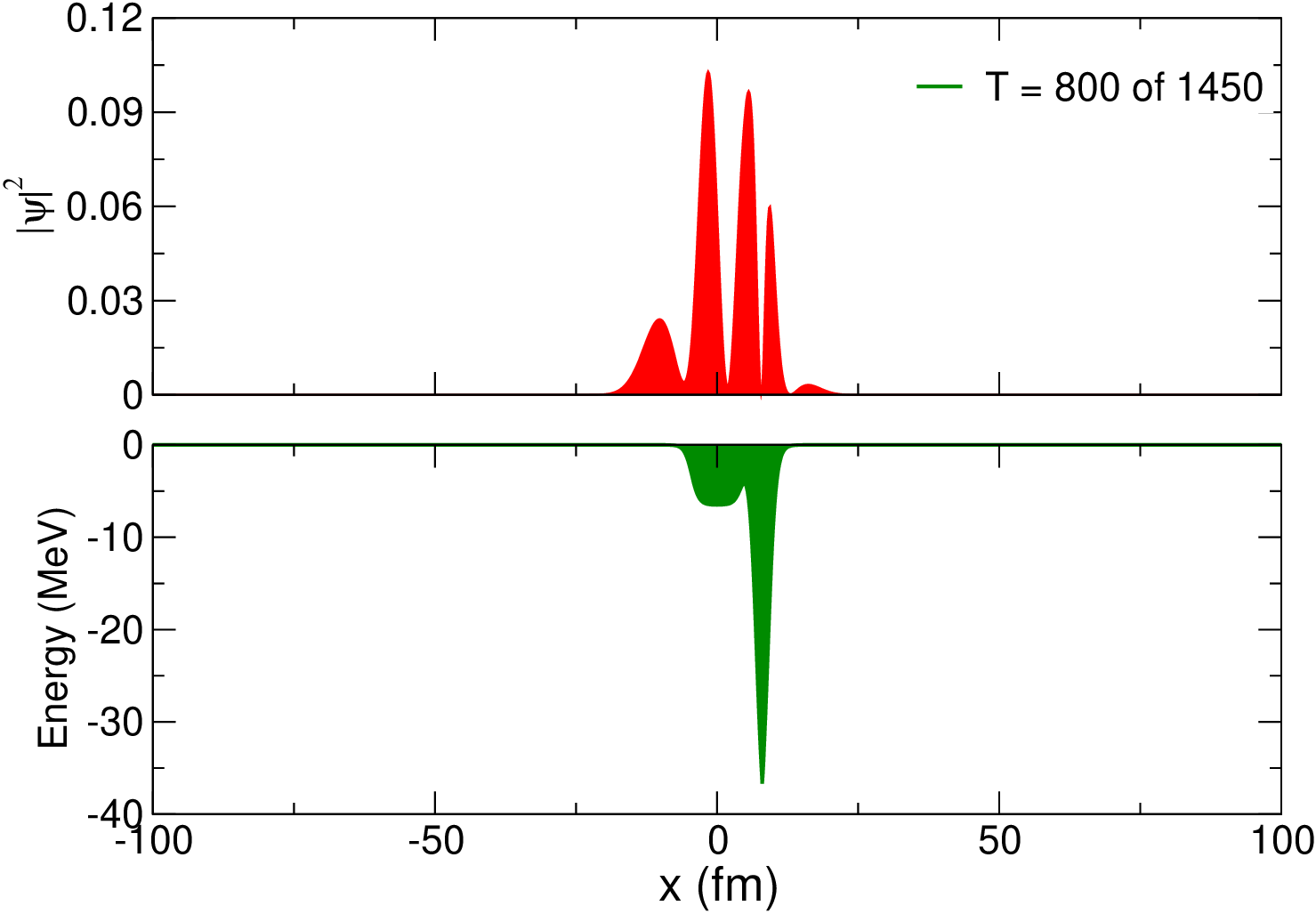}}
    \vspace{0.1cm}
    \subfigure[]{\includegraphics[scale=.21]{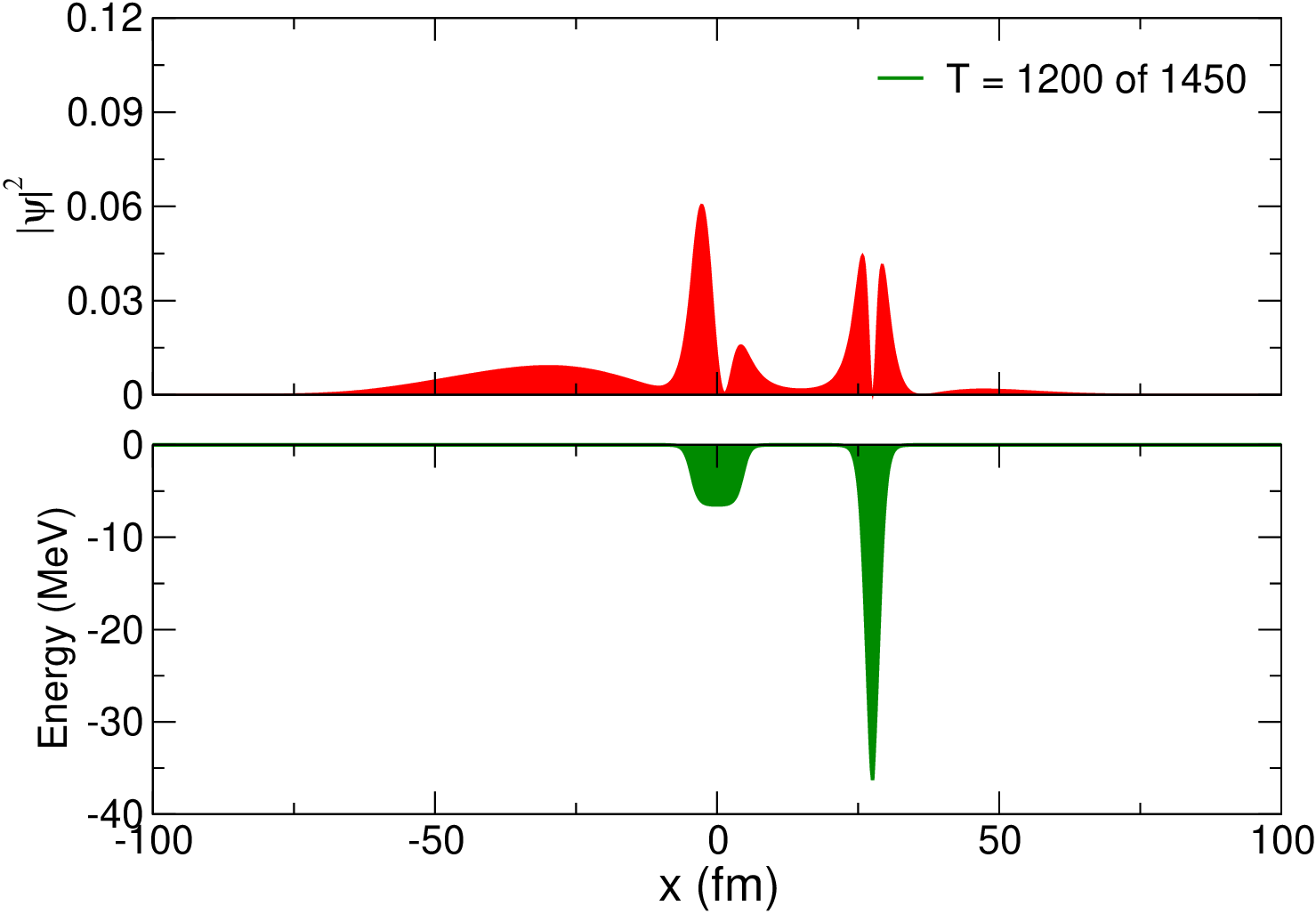}}
    \subfigure[]{\includegraphics[scale=.21]{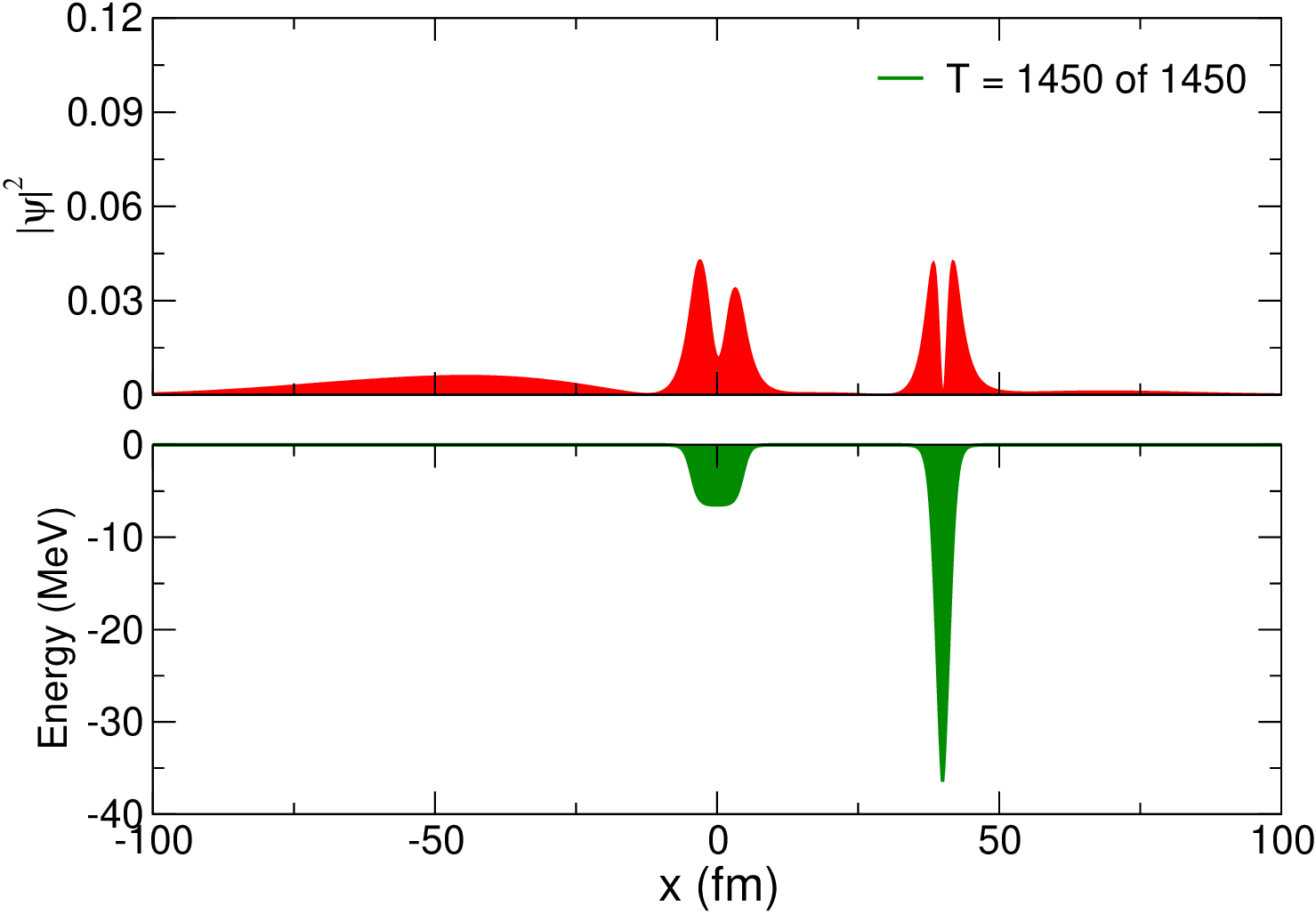}}
    \caption{Model case C. 
      Upper panels: the squared total wavefunction at different times with exact method.
      Lower panels: the target and projectile potential wells. 
      The elapsed time for each frame is indicated in the legends.}
    \label{fig_caseC_wf}
\end{figure}

The evolution of the exact wavefunction during the collision is presented in figure \ref{fig_caseC_wf},
where in each frame the central panel shows the exact wavefunction 
and the lower displays the target and projectiles potentials.
The initial wavefunction exhibits a node, because the initial state corresponds to the first excited state of the well. At the end of the time evolution (Fig.~\ref{fig_caseC_wf}d) the component related to the target is 
asymmetric, and so we expect a probability to de-excite the target to its ground state.
 One can also clearly identify a continuum component (in the region around $-50$~fm in Fig.~\ref{fig_caseC_wf}(d), 
and the presence of transfer to the projectile excited state because  the wavefunction component which remains bound to that well presents a node.

\begin{figure}[!hbt]
\begin{center}
\includegraphics[scale=.5]{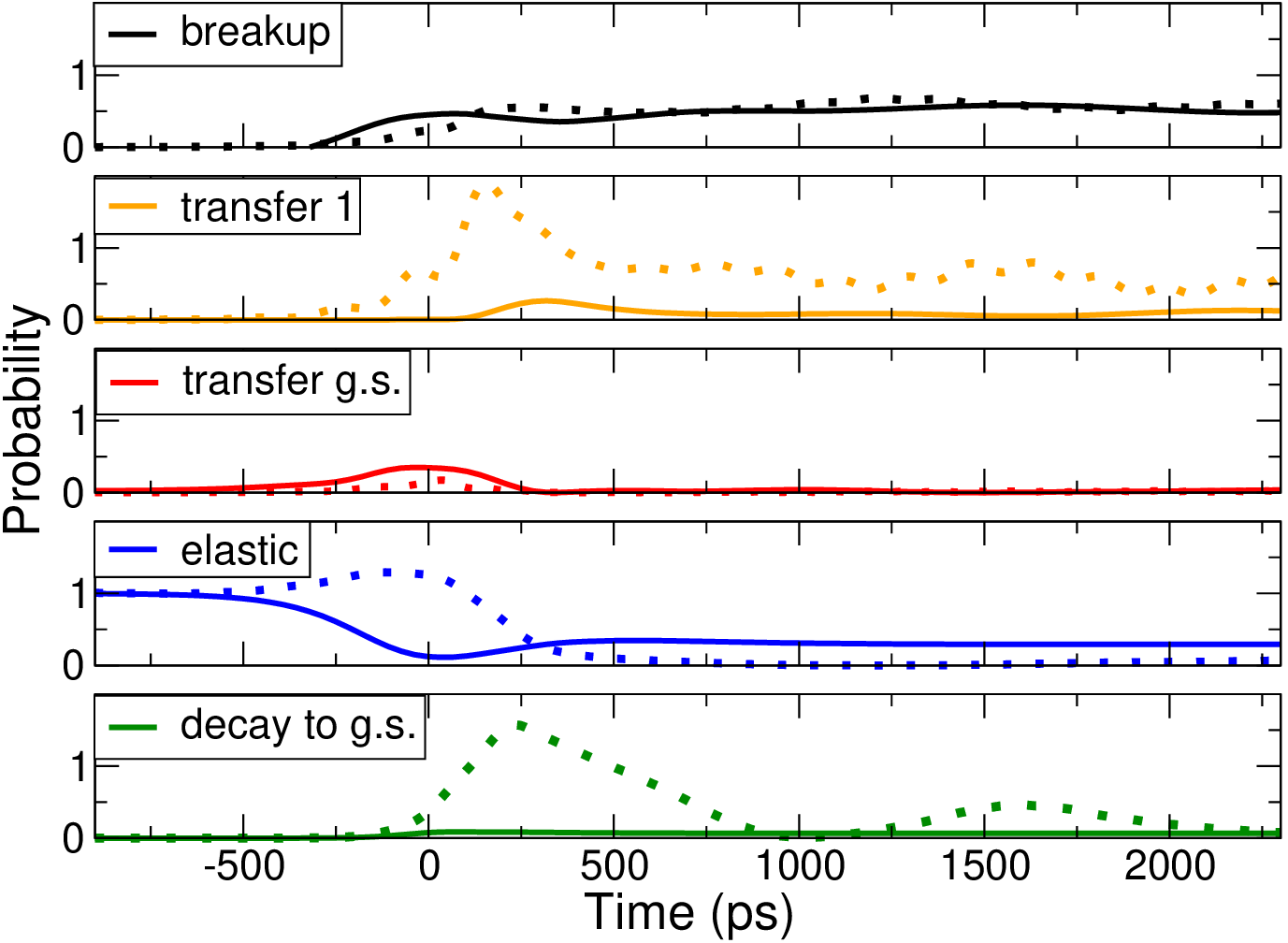}
\caption{Evolution of probability as a function of time for the model case C. 
  Each panel corresponds to a reaction channel indicated in the corresponding legend.
  Solid lines represent the exact results, 
  dotted lines correspond to the results obtained with coupled-channels calculation CC$^{(3)}$
  which are reported in table \ref{tab_case-C}.}
\label{fig_problVStime_case-C}
\end{center}
\end{figure}

The results for this case are reported in table \ref{tab_case-C}, and correspond to the following calculations:
\begin{figure}[!hbt]
\begin{center}
\includegraphics[scale=.35]{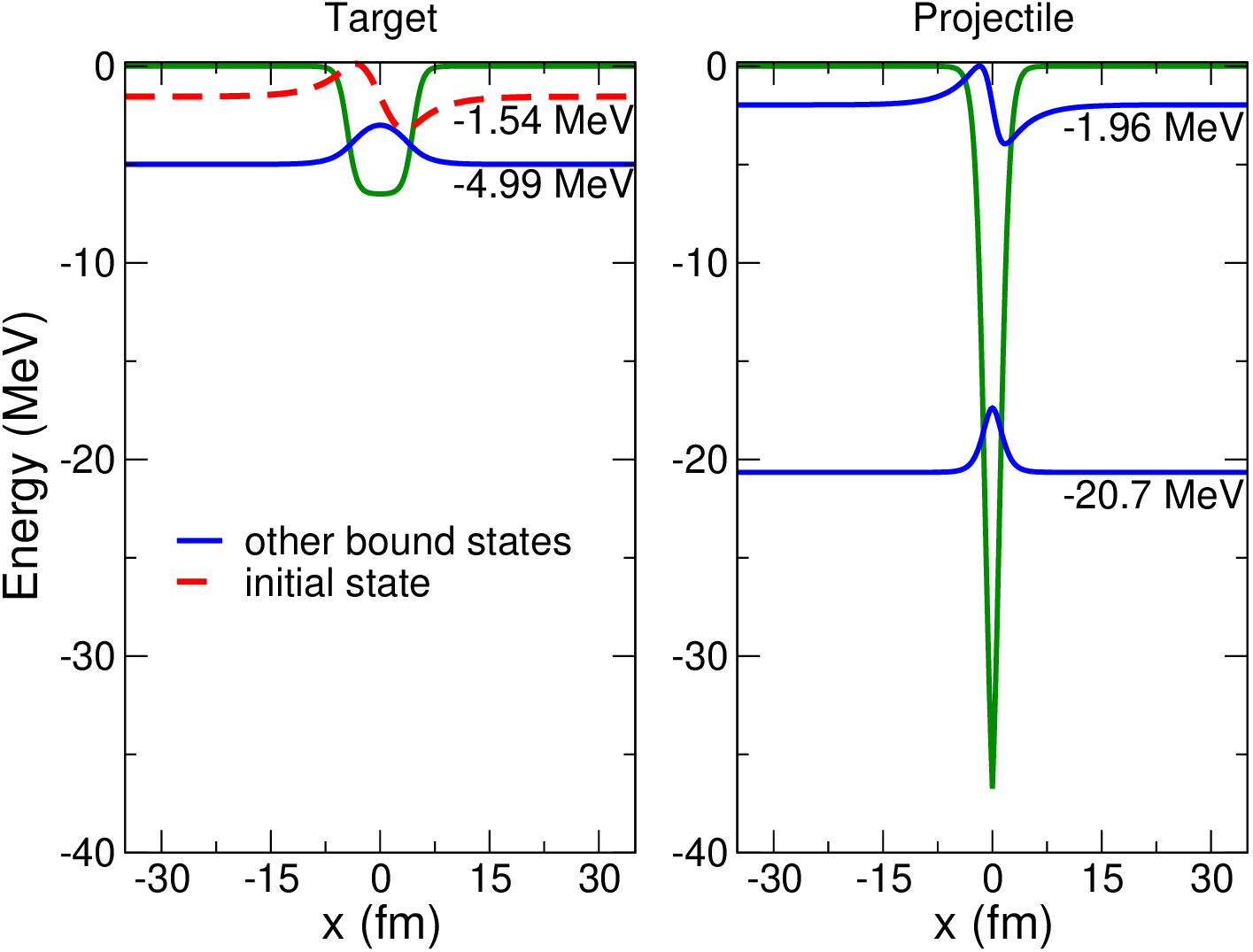}
\caption{Target (left) and projectile (right) Woods-Saxon potential and corresponding bound states for the model case C. The initial state, in this case the target excited state, is the dashed red curve.}
\label{pot:caseC}
\end{center}
\end{figure}
%
\begin{table}[h!]
\begin{center}
  \begin{tabular}{ l | c | c | c | c | c }
                  & Exact     & CC$^{(1)}$ & CC$^{(2)}$ & CC$^{(3)}$ & CC$^{(4)}$\\
    \hline
 Decay to g.s. & 6.7$\%$   & 0.85$\%$ & 23.5$\%$     & 16$\%$   & 3$\%$\\ 
    Elastic       & 29$\%$    &  99.06$\%$ &  44.8$\%$    & 10$\%$   & 84$\%$\\ 
    \hline
    Transfer g.s. & 1.2$\%$   &  0.03$\%$ & -    & 0.6$\%$  & 0.04$\%$\\
    Transfer 1st    & 15.6$\%$  &  0.06$\%$ & -    & 36$\%$   & 0.2$\%$\\
    \hline
    Breakup       & 47.5$\%$  &   - & 31.7$\%$      & 44$\%$   & 20$\%$\\
    \hline
  \end{tabular}
  \caption{Final probabilities for the model case C
    obtained with the exact and the coupled-channels methods.       
    Note that CC$^{(3)}$ and CC$^{(4)}$ total probability exceeds 100$\%$, as expected.}
  \label{tab_case-C}
\end{center}
\end{table}
\begin{itemize}
\item \textbf{Exact:} Calculation performed by solving numerically the time-dependent Schr\"odinger equation
                        (\ref{Schr\"odinger_equation_time_dependent}).
\item \textbf{CC$^{(1)}$:}  Coupled-channels calculation in which the target and projectile  bound states are included.
\item \textbf{CC$^{(2)}$:}  Coupled-channels calculation in which the target bound states and 10 target continuum states
                           within the $\left[-100;100\right]$~fm range    
                            are included.
\item \textbf{CC$^{(3)}$:}  Coupled-channels calculation using the target and projectile bound levels 
                          plus the first $10$ continuum pseudostates  of the target potential, obtained 
                          within a radius equal to the maximum radius of the grid.
                          It corresponds to an energy cutoff in the continuum of $0.3$ MeV.   
\item \textbf{CC$^{(4)}$:} Coupled-channels calculation using the target and projectile bound levels 
                          plus the first $10$ continuum pseudostates  of the projectile potential, obtained 
                          in a range equal to the maximum radius of the grid.
                          It corresponds to an energy cutoff in the continuum of $0.3$ MeV.   
\end{itemize}
As evident from \textbf{CC$^{(1)}$}, the inclusion of the transfer channel bound states is not enough and 
one needs the inclusion of continuum, thus indicating that this is not a one-step process in this case.
From the \textbf{CC$^{(2)}$} calculation we understand that the inclusion of the target continuum allows to reproduce reasonably well
the elastic and breakup channels, as well as the decay to the target ground state; but of course the transfer description is missing because the projectile basis is not included.
In this case in which all the reaction channels are relevant, both target and projectile bases including continuum states 
should be considered in our approximate calculation.
However in this case we have difficulties to apply the same method for the inclusion of continuum in coupled-channels calculation, 
as was done in the previous cases.
In case A the initial parameters were set in order to give 
a very small transfer probability to the projectile states,
thus the inclusion of target states was enough to reproduce the exact results.
In case B the breakup component did not escape quickly from the collision area, 
so at the end of the time evolution it was localized around the target 
in an interval that did not overlap with the projectile potential. 
This fact allowed us to construct two bases, for target and projectile respectively,
which did not overlap at the end of the calculation, thus restoring the probability unitarity. 
In the case under study, the breakup channel has a strong influence 
because it is travelling faster than the projectile,
and its component is not easily localizable within the  spatial grid.
What we propose here is to use continuum states which are defined over all the spatial range.
This is certainly closer to what the real continuum is (even neglecting the phaseshifts due to one of the wells), but surely will not lead to a unitary solution, 
because of overlap between bound states of a potential well and continuum pseudostates of the other.
By including few target continuum states (\textbf{CC$^{(3)}$}), e.g.\ $10$ pseudostates, 
the deviation from unitarity is not so large,
and we are still able to reproduce all the reaction channels.
The fact that the inclusion of projectile pseudostates (\textbf{CC$^{(4)}$})
does not reproduce the exact results tells us that the breakup component 
is mostly influenced by the target potential.
Since the projectile well is moving, 
the overlap between pseudostates and bound states is changing,
so we do not expect the coefficients associated to continuum  pseudostates to converge to a fixed value without oscillations.
In figure \ref{fig_problVStime_case-C} we show the time evolution of the probabilities computed with the \textbf{CC$^{(3)}$} calculation (dotted lines).
Each panel corresponds to a given reaction channel; 
from the lowest: decay to target ground-state (green), elastic scattering (blue), transfer to projectile ground state (red),
transfer to projectile first excited state (orange), and breakup (black).
The comparison with the ``exact'' result (solid lines) is presented. 
Note how the coupled-channels calculation clearly loses unitarity close to the turning point at $t = 0$~ps.
By applying this method we observe an acceptable description of the breakup channel probability by the coupled-channels calculation with continuum pseudostates, that oscillates around the expected exact value.

The situation presented in this case in which all the reaction channels are competing goes beyond the expected scope of applicability of standard coupled-channels methods. In fact, this situation resembles the case in which the distance  between the fragments following the projectile breakup ($|\textbf{r}|$) is comparable to the target-projectile distance ($|\textbf{R}|$). The Faddeev formalism should be better applied \cite{Faddeev}.

Anyhow, to obtain a proper description of the breakup we should use a set of continuum states which takes into account the effect of the two potential wells together. We have constructed a set of BOX pseudostates for the system including both potentials with Hamiltonian
\begin{equation}
  {\cal H}(x,t_f) = -\frac{\hbar^2}{2\mu}\frac{\partial ^2}{\partial x^2} + V_T(x) +  V_P(x-x_p(t_f)),
  \label{Hamiltonian_t_f}
\end{equation}
using the whole spatial grid. We have then projected the final exact wavefunction onto this set of continuum states to evaluate how they describe the breakup channel probability. In table \ref{tab_case-C_TP} we show this probability varying the number of pseudostates included in the basis. Increasing N, the total probability tends to the exact result. We need the contribution of a large number of states to get the proper result.

\begin{table}[h!]
\begin{center}
  \begin{tabular}{  c | c  }
     N     & Breakup probability \\
    \hline
    30 & 17.8 \% \\
    50 & 41.6 \%\\
    55 & 43.5 \%\\
    60 & 46.7 \%\\
    62   & 47.3 \%\\  
    63  &  47.5 \% \\
    \hline
     \textbf{Exact result} & \textbf{47.5\% }\\
     \hline
  \end{tabular}
  \caption{Breakup probabilities for the model case C 
    obtained with the projection of the final exact wavefunction onto N continuum pseudostates calculated taking into account both potential wells in the final configuration.}
  \label{tab_case-C_TP}
\end{center}
\end{table}
In a time-dependent model, to include the continuum of all the constituents at the same time one should compute a complete basis (including bound and continuum pseudostates) for the total system at each time step. This follows the so-called ``two-center'' method, widely used in atomic physics and recently applied for nuclear physics calculations \cite{DIAZTORRES2005373,DT08,DIAZTORRES2018381}.

It is worth mentioning here that a difference respect to the usual calculation is that in our framework we only included real potentials: the use of an imaginary component can account for the fusion and other excluded channels and could solve the non-unitary issue of the present formalism. This solution could also account for the neglected target recoil in our model. In our model the trajectory is fixed and the target recoil is neglected: so the energy is not conserved in these processes. However, this is not expected to affect the result, when the asymptotic energy is much higher than the excitation one. In the situation shown in the last case (section \ref{caseC}),  the excitation energy might be a significant fraction of the available kinetic energy or even be of the same order. This resembles the limitations encountered in the eikonal model \cite{glauber} (which is based on the assumption of straight projectile trajectories) when it is applied to study reactions at low beam energy \cite{AZV97,HC17,HC18}. 
Even if we used real potentials, the main conclusions on the role of continuum in the reaction mechanism remain unchanged. The description of a continuum using a basis which takes into account the phase shift induced by only one potential well, and, more in general, the choice of the continuum which describes only a subsystem of the reaction constituents, are not good approximations in the case of dominant breakup channel. 

\section{Conclusions \label{conclusion}}
We studied processes involving both bound and weakly-bound systems, and all the possible channels of a direct reaction  within a simple one-dimensional problem consisting in the scattering of a particle, initially bound in a potential well (the {\it target}), by another moving well (the {\it projectile}). An appealing feature of this  problem is that it can be solved ``exactly'' using the time-dependent Schrodinger equation, thus providing a robust benchmark for approximate methods, such as the popular coupled-channels formalism employed in atomic, molecular and nuclear physics.
In this framework, we can investigate the role of continuum, which is found to be relevant even for situations in which the considered particle is initially in a well-bound state. In particular, we address the issue of the inclusion of a discretized continuum in a coupled-channels time-dependent  calculation. 
Continuum waves are treated in our computational model in ``mathematical'' representations of non-orthogonal sets of normalizable wavefunctions that reflect the phase shifts induced by each potential well.
We observe that when the breakup channel is almost negligible respect the other direct reaction channels, like our case A, this representation of  the continuum works well.
When the breakup clearly dominates or is competing with the other channels, like in cases B and C, the inclusion of a set of continuum pseudostates associated to only one of the potential wells is not enough to properly describe the reaction. However, the spatial limitations to continuum might give only a partial description of the process. So, the inclusion of continuum in computational models is always an artifact, and in many cases the results depend on how the continuum is constructed, as clearly shown in 1D. Therefore, the discretized continuum defined taking into account both potential wells on the whole range should be the best approximation.  To take into account this configuration in a time-dependent picture in the present formalism is however not possible, so we suggest the solution of the ``two-center'' method \cite{DIAZTORRES2005373,DT08,DIAZTORRES2018381}. The natural extension of this study would be then the development of a continuum-discretized coupled-channels calculation within the ``two-center'' formalism, to be compared to the exact solution of the time-dependent Schr\"odinger equation.\\
The present framework could also be applied to the description of direct reactions involving two-neutron halo nuclei, e.g.\ to study the role of pairing between the two valence particles. First results for the exact solution of this process could be found in \cite{Moschini_thesis:2017,VMHM2015,Vitturi_2018}; alternative 1D time-dependent approaches are described in \cite{BD15}.

\section*{Acknowledgments}
We thank Professor Alexis Diaz-Torres for a careful reading of the paper, helpful discussions, and comments.
LM is grateful to Università degli Studi di Padova and Universidad de Sevilla, where the majority of this project has been developed under INFN-Padova financial support.

\section*{Bibliography}

\begin{thebibliography}{10}

\bibitem{Tanihata3}
I.~Tanihata, H.~Hamagaki, O.~Hashimoto, Y.~Shida, N.~Yoshikawa, K.~Sugimoto,
  O.~Yamakawa, T.~Kobayashi, and N.~Takahashi {\em Phys.\ Rev.\ Lett.},
  vol.~55, p.~2676, 1985.

\bibitem{Tanihata4}
I.~Tanihata, T.~Kobayashi, O.~Yamakawa, S.~Shimoura, K.~Ekuni, K.~Sugimoto,
  N.~Takahashi, T.~Shimoda, and H.~Sato {\em Phys.\ Lett.\ B}, vol.~206,
  pp.~592--596, 1988.

\bibitem{Nilsson1}
S.~G. Nilsson {\em K.\ Dan.\ Vidensk.\ Selsk.\ Mat.\ Fys.\ Medd.}, vol.~29,
  no.~16, 1955.

\bibitem{Nilsson2}
B.~R. Mottelson and S.~G. Nilsson {\em Mat.\ Fys.\ Skr.\ Dan.\ Vid.\ Selsk.},
  vol.~1, no.~8, 1959.

\bibitem{Heyde}
K.~Heyde, {\em Basic Ideas and Concepts in Nuclear Physics: An Introductory
  Approach}.
\newblock IOP Publishing, 1999.

\bibitem{Tanihata}
I.~Tanihata {\em J.\ Phys.\ G: Nucl.\ Part.\ Phys.}, vol.~22, p.~157, 1996.

\bibitem{Tanihata2}
I.~Tanihata, H.~Savajols, and R.~Kanungo {\em Prog.\ Part.\ Nucl.\ Phys.},
  vol.~68, pp.~215--313, 2013.

\bibitem{TANIHATA1995505}
I.~Tanihata, ``Nuclear structure studies from reaction induced by radioactive
  nuclear beams,'' {\em Progress in Particle and Nuclear Physics}, vol.~35,
  pp.~505 -- 573, 1995.

\bibitem{Satchler}
G.~R. Satchler, {\em Introduction to Nuclear Reactions}.
\newblock The Macmillan Press LTD, 1980.

\bibitem{CDCC1}
G.~Rawitscher {\em Phys.\ Rev.\ C}, vol.~9, p.~2210, 1974.

\bibitem{AUSTERN1987125}
N.~Austern, Y.~Iseri, M.~Kamimura, M.~Kawai, G.~Rawitscher, and M.~Yahiro,
  ``Continuum-discretized coupled-channels calculations for three-body models
  of deuteron-nucleus reactions,'' {\em Physics Reports}, vol.~154, no.~3,
  pp.~125 -- 204, 1987.

\bibitem{thompson_nunes_2009}
I.~J. Thompson and F.~M. Nunes, {\em Nuclear Reactions for Astrophysics:
  Principles, Calculation and Applications of Low-Energy Reactions}.
\newblock Cambridge University Press, 2009.

\bibitem{YOMM12}
M.~Yahiro, K.~Ogata, T.~Matsumoto, and K.~Minomo, ``{The continuum discretized
  coupled-channels method and its applications},'' {\em Progress of Theoretical
  and Experimental Physics}, vol.~2012, 09 2012.
\newblock 01A206.

\bibitem{cdcc-lim1}
T.~Sawada and K.~Thushima {\em Progr.\ Theoret.\ Phys.}, vol.~76, p.~440, 1986.

\bibitem{cdcc-lim2}
Z.~C. Kuruo\u{g}lu {\em Phys.\ Rev.\ C}, vol.~43, p.~1061, 1991.

\bibitem{cdcc-lim3}
N.~J. Upadhyay, A.~Deltuva, and F.~M. Nunes {\em Phys.\ Rev.\ C}, vol.~85,
  p.~054621, 2012.

\bibitem{Esbensen}
H.~Esbensen, R.~A. Broglia, and A.~Winther {\em Ann. Phys.}, vol.~146,
  pp.~149--173, 1983.

\bibitem{numerov}
C.~A. Moyer {\em Comp.\ Sci.\ Eng.}, vol.~8, pp.~32--40, 2006.

\bibitem{MPV2016}
L.~Moschini, F.~P{\'{e}}rez-Bernal, and A.~Vitturi, ``Bound and unbound nuclear
  systems at the drip lines: a one-dimensional model,'' {\em Journal of Physics
  G: Nuclear and Particle Physics}, vol.~43, p.~045112, mar 2016.

\bibitem{Esbensen-ref7}
P.~Bonche, S.~Koonin, and J.~W. Negele {\em Phys.\ Rev.\ C}, vol.~13, p.~1226,
  1976.

\bibitem{Moschini_thesis:2017}
L.~Moschini, {\em Ph.D. Thesis}, 2017.

\bibitem{Esbensen-ref4}
R.~A. Broglia and A.~Winther {\em Nucl.\ Phys.\ A}, vol.~182, p.~112, 1972.

\bibitem{Esbensen-ref5}
R.~A. Broglia and A.~Winther {\em Phys.\ Rep.\ C}, vol.~4, p.~154, 1972.

\bibitem{Esbensen-ref6}
K.~Dietrich and K.~Hara {\em Nucl.\ Phys.\ A}, vol.~211, p.~349, 1973.

\bibitem{Faddeev}
L.~D. Faddeev {\em Zh.\ Eksp.\ Theor.\ Fiz.}, vol.~39, p.~1459, 1960.

\bibitem{M2014}
L.~Moschini {\em Journal of Physics: Conference Series}, vol.~566, no.~1,
  p.~012027, 2014.

\bibitem{VM2015}
A.~Vitturi and L.~Moschini, ``Structure and dynamics of weakly-bound systems: a
  one-dimensional model,'' {\em Journal of Physics: Conference Series},
  vol.~590, no.~1, p.~012007, 2015.

\bibitem{MVM2016}
L.~Moschini, A.~Vitturi, and A.~Moro, ``Direct reactions: A one dimensional
  toy-model,'' in {\em Basic Concepts in Nuclear Physics: Theory, Experiments
  and Applications} (J.-E. Garc{\'i}a-Ramos, C.~E. Alonso, M.~V. Andr{\'e}s,
  and F.~P{\'e}rez-Bernal, eds.), (Cham), pp.~181--183, Springer International
  Publishing, 2016.

\bibitem{MVA2018}
L.~Moschini, A.~Vitturi, and A.~Moro, ``Direct reactions of weakly-bound nuclei
  within a one dimensional model,'' {\em Journal of Physics: Conference
  Series}, vol.~981, no.~1, p.~012004, 2018.

\bibitem{FB12}
F.~Flavigny, A.~Obertelli, A.~Bonaccorso, G.~F. Grinyer, C.~Louchart,
  L.~Nalpas, and A.~Signoracci, ``Nonsudden limits of heavy-ion induced
  knockout reactions,'' {\em Phys. Rev. Lett.}, vol.~108, p.~252501, Jun 2012.

\bibitem{MCNT02}
A.~M. Moro, R.~Crespo, F.~Nunes, and I.~J. Thompson, ``${}^{8}\mathrm{B}$
  breakup in elastic and transfer reactions,'' {\em Phys. Rev. C}, vol.~66,
  p.~024612, Aug 2002.

\bibitem{LCAGQG09}
J.~Lubian, T.~Correa, E.~F. Aguilera, L.~F. Canto, A.~Gomez-Camacho, E.~M.
  Quiroz, and P.~R.~S. Gomes, ``Effects of breakup couplings on
  $^{8}\mathrm{B}+^{58}\mathrm{Ni}$ elastic scattering,'' {\em Phys. Rev. C},
  vol.~79, p.~064605, Jun 2009.

\bibitem{KAKR09}
N.~Keeley, N.~Alamanos, K.~Kemper, and K.~Rusek, ``Elastic scattering and
  reactions of light exotic beams,'' {\em Progress in Particle and Nuclear
  Physics}, vol.~63, no.~2, pp.~396 -- 447, 2009.

\bibitem{HVDL00}
K.~Hagino, A.~Vitturi, C.~H. Dasso, and S.~M. Lenzi, ``Role of breakup
  processes in fusion enhancement of drip-line nuclei at energies below the
  coulomb barrier,'' {\em Phys. Rev. C}, vol.~61, p.~037602, Feb 2000.

\bibitem{DT02}
A.~Diaz-Torres and I.~J. Thompson, ``Effect of continuum couplings in fusion of
  halo ${}^{11}\mathrm{Be}$ on ${}^{208}\mathrm{Pb}$ around the coulomb
  barrier,'' {\em Phys. Rev. C}, vol.~65, p.~024606, Jan 2002.

\bibitem{CLGH09}
L.~F. Canto, J.~Lubian, P.~R.~S. Gomes, and M.~S. Hussein,
  ``Continuum-continuum coupling and polarization potentials for weakly bound
  systems,'' {\em Phys. Rev. C}, vol.~80, p.~047601, Oct 2009.

\bibitem{NT99}
F.~M. Nunes and I.~J. Thompson, ``Multistep effects in sub-coulomb breakup,''
  {\em Phys. Rev. C}, vol.~59, pp.~2652--2659, May 1999.

\bibitem{LM19}
J.~Lei and A.~M. Moro, ``Unraveling the reaction mechanisms leading to partial
  fusion of weakly bound nuclei,'' {\em Phys. Rev. Lett.}, vol.~123, p.~232501,
  Dec 2019.

\bibitem{DMCNF}
A.~Deltuva, A.~M. Moro, E.~Cravo, F.~M. Nunes, and A.~C. Fonseca {\em Phys.\
  Rev.\ C}, vol.~79, p.~064602, 2009.

\bibitem{DIAZTORRES2005373}
A.~Diaz-Torres and W.~Scheid, ``Two center shell model with woods–saxon
  potentials: Adiabatic and diabatic states in fusion,'' {\em Nuclear Physics
  A}, vol.~757, no.~3, pp.~373 -- 389, 2005.

\bibitem{DT08}
A.~Diaz-Torres, ``Solving the two-center nuclear shell-model problem with
  arbitrarily oriented deformed potentials,'' {\em Phys. Rev. Lett.}, vol.~101,
  p.~122501, Sep 2008.

\bibitem{DIAZTORRES2018381}
A.~Diaz-Torres, ``Owl: A code for the two-center shell model with spherical
  woods–saxon potentials,'' {\em Computer Physics Communications}, vol.~224,
  pp.~381 -- 386, 2018.

\bibitem{glauber}
R.~Glauber, {\em Lectures in Theoretical Physics}, vol.~1.
\newblock New York: Interscience, 1959.

\bibitem{AZV97}
C.~E. Aguiar, F.~Zardi, and A.~Vitturi, ``Low-energy extensions of the eikonal
  approximation to heavy-ion scattering,'' {\em Phys. Rev. C}, vol.~56,
  pp.~1511--1515, Sep 1997.

\bibitem{HC17}
C.~Hebborn and P.~Capel, ``Analysis of corrections to the eikonal
  approximation,'' {\em Phys. Rev. C}, vol.~96, p.~054607, Nov 2017.

\bibitem{HC18}
C.~Hebborn and P.~Capel, ``Low-energy corrections to the eikonal description of
  elastic scattering and breakup of one-neutron halo nuclei in
  nuclear-dominated reactions,'' {\em Phys. Rev. C}, vol.~98, p.~044610, Oct
  2018.

\bibitem{VMHM2015}
A.~Vitturi, L.~Moschini, K.~Hagino, and A.~M. Moro {\em AIP Conference
  Proceedings}, vol.~1681, p.~060001, 2015.

\bibitem{Vitturi_2018}
A.~Vitturi and L.~Moschini, ``Interplay of break-up and transfer processes in
  reactions involving weakly-bound systems,'' {\em Journal of Physics:
  Conference Series}, vol.~966, p.~012045, feb 2018.

\bibitem{BD15}
M.~Boselli and A.~Diaz-Torres, ``Quantifying low-energy fusion dynamics of
  weakly bound nuclei from a time-dependent quantum perspective,'' {\em Phys.
  Rev. C}, vol.~92, p.~044610, Oct 2015.

\end{thebibliography}

\end{document}